\RequirePackage{ifpdf}
\documentclass[a4paper,11pt]{article}
\pdfoutput=1 
\usepackage{jheppub} 
\usepackage[T1]{fontenc} 
\usepackage{ifpdf} 
\usepackage{accents} 

\makeatletter
\renewcommand\@fpheader{} 
\renewcommand\@journal{}
\makeatother

\title{
The Two-Loop Master Integrals for $q \bar{q} \to V V$
}

\preprint{{ZU-TH 16/14, LPN14-065, MITP/14-021}}

\author[a]{Thomas Gehrmann,}
\author[b]{Andreas von Manteuffel,}
\author[a]{Lorenzo Tancredi,}
\author[a]{Erich Weihs}

\affiliation[a]{
  Physik-Institut, 
  Universit\"at Z\"urich, Wintherturerstrasse 190,
  CH-8057~Z\"urich, Switzerland}

  \affiliation[b]{
  PRISMA Cluster of Excellence \& Institute of Physics,
  Johannes Gutenberg University,\\
  55099~Mainz, Germany}

\emailAdd{thomas.gehrmann@uzh.ch}
\emailAdd{manteuffel@uni-mainz.de}
\emailAdd{tancredi@physik.uzh.ch}
\emailAdd{erich.weihs@physik.uzh.ch}

\keywords{QCD, Feynman integrals, Polylogarithms, NLO and NNLO Calculations}

\abstract{We compute the full set of two-loop Feynman integrals appearing in 
massless two-loop four-point functions with two off-shell legs with the same invariant mass. 
These integrals allow to determine the two-loop corrections to the 
amplitudes for vector boson pair production 
at hadron colliders, $q \bar{q} \to V V$, and thus to compute this process to 
next-to-next-to-leading order accuracy in QCD. 
The master integrals are derived using the method of differential equations, employing a 
canonical basis for the integrals. We obtain analytical results for all integrals, expressed in 
terms of multiple polylogarithms.
We optimize our results for numerical evaluation
by employing functions which are real valued for physical scattering kinematics
and allow for an immediate power series expansion.
}

\ifpdf
\fi

\def\Q2{\left(Q^{2}\right)}

\def\Li{\hbox{Li}}

\newcommand{\be}{\begin{equation}}
\newcommand{\ee}{\end{equation}}
\newcommand{\bea}{\begin{eqnarray}}
\newcommand{\eea}{\end{eqnarray}}
\newcommand{\vm}{\vec{m}}
\newcommand{\vf}{\vec{f}}

\newcommand{\C}{\mathcal{C}}

\newcommand{\Z}{\mathbb{Z}}

\newcommand{\labbel}[1] { \label{#1} } 

\newcommand\thickbar[1]{\accentset{\rule{.4em}{.8pt}}{#1}}
\newcommand{\topoid}[1]{{\scalebox{0.7}[0.7]{$\mathrm{#1}$}}}

\def\bsp#1\esp{\begin{split}#1\end{split}}

\newcommand{\bubbleNLO}[1]{
\mbox{\parbox{2.5cm}{\hspace{-0.1cm}
\begin{picture}(2,1)
\put(1.0,0.97){\circle*{0.15}}
\put(1.0,0.5){\circle*{0.15}}
\put(0.3,0.5){\vector(1,0){0.1}}
\put(0.5,0.5){\line(1,0){1}}
\linethickness{0.50mm}
\put(0,0.5){\line(1,0){0.5}}
\put(1.5,0.5){\line(1,0){0.5}}
\thinlines
\put(1,0.5){\circle{1}}
\put(0.25,0.7){\makebox(0,0)[b]{$#1$}}
\end{picture}
}}
\hfill}

\newcommand{\doublebubble}[1]{
\mbox{\parbox{3.5cm}{\hspace{-0.1cm}
\begin{picture}(3,1)
\put(1.0,0.97){\circle*{0.15}}
\put(2.0,0.97){\circle*{0.15}}
\put(0.3,0.5){\vector(1,0){0.1}}
\linethickness{0.50mm}
\put(0,0.5){\line(1,0){0.5}}
\put(2.5,0.5){\line(1,0){0.5}}
\thinlines
\put(1,0.5){\circle{1}}
\put(2,0.5){\circle{1}}
\put(0.25,0.7){\makebox(0,0)[b]{$#1$}}
\end{picture}
}}
\hfill}

\newcommand{\doublebubblex}[3]{
\mbox{\parbox{3.5cm}{\hspace{-0.1cm}
\begin{picture}(3,1)
\put(1.0,0.97){\circle*{0.15}}
\put(2.0,0.97){\circle*{0.15}}
\linethickness{0.50mm}
\put(0,0.5){\line(1,0){0.5}}
\put(2.5,0.5){\line(1,0){0.5}}
\put(1.5,0.5){\line(0,1){0.6}}
\thinlines
\put(0.3,0.5){\vector(1,0){0.1}}
\put(2.75,0.5){\vector(1,0){0.1}}
\put(1.5,1.0){\vector(0,1){0.1}}
\put(0,0.5){\line(1,0){0.5}}
\put(2.5,0.5){\line(1,0){0.5}}
\put(1,0.5){\circle{1}}
\put(2,0.5){\circle{1}}
\put(0.25,0.7){\makebox(0,0)[b]{$#1$}}
\put(2.75,0.7){\makebox(0,0)[b]{$#2$}}
\put(1.5,1.25){\makebox(0,0)[b]{$#3$}}
\end{picture}
}}
\hfill}

\newcommand{\doublebubblexb}[3]{
\mbox{\parbox{3.5cm}{\hspace{-0.1cm}
\begin{picture}(3,1)
\put(1.0,0.97){\circle*{0.15}}
\put(2.0,0.97){\circle*{0.15}}
\linethickness{0.50mm}
\put(0,0.5){\line(1,0){0.5}}
\put(2.5,0.5){\line(1,0){0.5}}
\put(1.5,0.5){\line(0,1){0.6}}
\thinlines
\put(0.3,0.5){\vector(1,0){0.1}}
\put(2.75,0.5){\vector(1,0){0.1}}
\put(1.5,1.0){\vector(0,-1){0.1}}
\put(0,0.5){\line(1,0){0.5}}
\put(2.5,0.5){\line(1,0){0.5}}
\put(1,0.5){\circle{1}}
\put(2,0.5){\circle{1}}
\put(0.25,0.7){\makebox(0,0)[b]{$#1$}}
\put(2.75,0.7){\makebox(0,0)[b]{$#2$}}
\put(1.5,1.25){\makebox(0,0)[b]{$#3$}}
\end{picture}
}}
\hfill}

\newcommand{\doublebubblexc}[3]{
\mbox{\parbox{3.5cm}{\hspace{-0.1cm}
\begin{picture}(3,1)
\put(1.0,0.97){\circle*{0.15}}
\put(2.0,0.97){\circle*{0.15}}
\linethickness{0.50mm}
\put(0,0.5){\line(1,0){0.5}}
\put(2.5,0.5){\line(1,0){0.5}}
\put(1.5,0.5){\line(0,1){0.6}}
\thinlines
\put(0.3,0.5){\vector(-1,0){0.1}}
\put(2.75,0.5){\vector(1,0){0.1}}
\put(1.5,1.0){\vector(0,-1){0.1}}
\put(0,0.5){\line(1,0){0.5}}
\put(2.5,0.5){\line(1,0){0.5}}
\put(1,0.5){\circle{1}}
\put(2,0.5){\circle{1}}
\put(0.25,0.7){\makebox(0,0)[b]{$#1$}}
\put(2.75,0.7){\makebox(0,0)[b]{$#2$}}
\put(1.5,1.25){\makebox(0,0)[b]{$#3$}}
\end{picture}
}}
\hfill}

\newcommand{\trianglebubble}[3]{
\mbox{\parbox{3cm}{\hspace{0.0cm}
\begin{picture}(2.5,1.4)
\put(1.25,1.42){\circle*{0.15}}
\put(0.3,0.7){\vector(1,0){0.1}}
\put(1.9,0.2){\vector(1,0){0.1}}
\put(1.9,1.2){\vector(1,0){0.1}}
\linethickness{0.50mm}
\put(0,0.7){\line(1,0){0.5}}
\put(1.5,1.2){\line(1,0){0.5}}
\put(1.5,0.2){\line(1,0){0.5}}
\thinlines
\put(0.5,0.7){\line(2,-1){1}}
\put(0.5,0.7){\line(1,1){0.5}}
\put(1.5,1.2){\line(0,-1){1}}
\put(1.25,1.2){\circle{0.5}}
\put(0.25,0.9){\makebox(0,0)[b]{$#1$}}
\put(2.05,1.2){\makebox(0,0)[l]{$#2$}}
\put(2.05,0.2){\makebox(0,0)[l]{$#3$}}
\end{picture}
}}
\hfill}

\newcommand{\trianglebubbleb}[3]{
\mbox{\parbox{3cm}{\hspace{0.0cm}
\begin{picture}(2.5,1.4)
\put(1.25,1.42){\circle*{0.15}}
\put(0.3,0.7){\vector(-1,0){0.1}}
\put(1.9,0.2){\vector(1,0){0.1}}
\put(1.9,1.2){\vector(-1,0){0.1}}
\linethickness{0.50mm}
\put(0,0.7){\line(1,0){0.5}}
\put(1.5,1.2){\line(1,0){0.5}}
\put(1.5,0.2){\line(1,0){0.5}}
\thinlines
\put(0.5,0.7){\line(2,-1){1}}
\put(0.5,0.7){\line(1,1){0.5}}
\put(1.5,1.2){\line(0,-1){1}}
\put(1.25,1.2){\circle{0.5}}
\put(0.25,0.9){\makebox(0,0)[b]{$#1$}}
\put(2.05,1.2){\makebox(0,0)[l]{$#2$}}
\put(2.05,0.2){\makebox(0,0)[l]{$#3$}}
\end{picture}
}}
\hfill}

\newcommand{\boxbubble}[4]{
\mbox{\parbox{3.5cm}{\hspace{-0.1cm}
\begin{picture}(3.5,1.4)
\put(0.7,0.2){\vector(1,0){0.1}}
\put(2.3,0.2){\vector(1,0){0.1}}
\put(2.3,1.2){\vector(1,0){0.1}}
\put(0.7,1.2){\vector(1,0){0.1}}
\put(2.25,0.95){\circle*{0.15}}
\put(0.5,0.2){\line(1,0){2.0}}
\put(0.5,1.2){\line(1,0){2.0}}
\put(1,0.2){\line(0,1){1}}
\put(2,0.7){\line(0,-1){0.5}}
\put(2,0.95){\circle{0.5}}
\linethickness{0.50mm}
\put(2.0,0.2){\line(1,0){0.5}}
\put(2.0,1.2){\line(1,0){0.5}}
\thinlines
\put(0.45,1.2){\makebox(0,0)[r]{$#1$}}
\put(0.45,0.2){\makebox(0,0)[r]{$#2$}}
\put(2.55,1.2){\makebox(0,0)[l]{$#3$}}
\put(2.55,0.2){\makebox(0,0)[l]{$#4$}}

\end{picture}
}} 
\hfill}

\newcommand{\triangleOne}[3]{
\mbox{\parbox{3cm}{\hspace{-0.1cm}
\begin{picture}(2.5,1.4)
\put(1.0,0.7){\circle*{0.15}}
\put(0.3,0.7){\vector(1,0){0.1}}
\put(1.7,0.2){\vector(1,0){0.1}}
\put(1.7,1.2){\vector(1,0){0.1}}
\linethickness{0.50mm}
\put(0,0.7){\line(1,0){0.5}}
\thinlines
\put(1,1.2){\line(0,-1){1}}
\put(1,1.2){\line(1,0){1}}
\put(1,0.2){\line(1,0){1}}
\put(1,0.7){\circle{1}}
\put(0.25,0.9){\makebox(0,0)[b]{$#1$}}
\put(2.05,1.2){\makebox(0,0)[l]{$#2$}}
\put(2.05,0.2){\makebox(0,0)[l]{$#3$}}
\end{picture}
}}
\hfill}

\newcommand{\triangleTwo}[3]{
\mbox{\parbox{3cm}{\hspace{-0.1cm}
\begin{picture}(2.5,1.4)
\put(1.0,0.7){\circle*{0.15}}
\put(0.3,0.7){\vector(1,0){0.1}}
\put(1.7,0.2){\vector(1,0){0.1}}
\put(1.7,1.2){\vector(1,0){0.1}}
\put(1,1.2){\line(0,-1){1}}
\put(1,1.2){\line(1,0){1}}
\linethickness{0.50mm}
\put(0,0.7){\line(1,0){0.5}}
\put(1,0.2){\line(1,0){1}}
\thinlines
\put(1,0.7){\circle{1}}
\put(0.25,0.9){\makebox(0,0)[b]{$#1$}}
\put(2.05,1.2){\makebox(0,0)[l]{$#2$}}
\put(2.05,0.2){\makebox(0,0)[l]{$#3$}}
\end{picture}
}}
\hfill}

\newcommand{\triangleTwox}[3]{
\mbox{\parbox{3cm}{\hspace{-0.1cm}
\begin{picture}(2.5,1.4)
\put(0.85,0.85){\circle*{0.15}}
\put(0.3,0.7){\vector(1,0){0.1}}
\put(1.7,0.2){\vector(1,0){0.1}}
\put(1.7,1.2){\vector(1,0){0.1}}
\put(0.5,1.2){\oval(1,1)[br]}
\put(1,1.2){\line(1,0){1}}
\linethickness{0.50mm}
\put(1,0.2){\line(1,0){1}}
\put(0,0.7){\line(1,0){0.5}}
\thinlines
\put(1,0.7){\circle{1}}
\put(0.25,0.9){\makebox(0,0)[b]{$#1$}}
\put(2.05,1.2){\makebox(0,0)[l]{$#2$}}
\put(2.05,0.2){\makebox(0,0)[l]{$#3$}}
\end{picture}
}}
\hfill}

\newcommand{\triangleThree}[3]{
\mbox{\parbox{3cm}{\hspace{-0.1cm}
\begin{picture}(2.5,1.4)
\put(0.3,0.7){\vector(1,0){0.1}}
\put(1.7,0.2){\vector(1,0){0.1}}
\put(1.7,1.2){\vector(1,0){0.1}}
\linethickness{0.50mm}
\put(0,0.7){\line(1,0){0.5}}
\put(1,1.2){\line(1,0){1}}
\put(1,0.2){\line(1,0){1}}
\thinlines
\put(1,1.2){\line(0,-1){1}}
\put(1,0.7){\circle{1}}
\put(0.25,0.9){\makebox(0,0)[b]{$#1$}}
\put(2.05,1.2){\makebox(0,0)[l]{$#2$}}
\put(2.05,0.2){\makebox(0,0)[l]{$#3$}}
\end{picture}
}}
\hfill}

\newcommand{\triangleThreedot}[3]{
\mbox{\parbox{3cm}{\hspace{-0.1cm}
\begin{picture}(2.5,1.4)
\put(0.3,0.7){\vector(1,0){0.1}}
\put(1.7,0.2){\vector(1,0){0.1}}
\put(1.7,1.2){\vector(1,0){0.1}}
\linethickness{0.50mm}
\put(0,0.7){\line(1,0){0.5}}
\put(1,1.2){\line(1,0){1}}
\put(1,0.2){\line(1,0){1}}
\thinlines
\put(1,1.2){\line(0,-1){1}}
\put(1.0,0.7){\circle*{0.15}}
\put(1,0.7){\circle{1}}
\put(0.25,0.9){\makebox(0,0)[b]{$#1$}}
\put(2.05,1.2){\makebox(0,0)[l]{$#2$}}
\put(2.05,0.2){\makebox(0,0)[l]{$#3$}}
\end{picture}
}}
\hfill}

\newcommand{\triangleThreex}[3]{
\mbox{\parbox{3cm}{\hspace{-0.1cm}
\begin{picture}(2.5,1.4)
\put(0.3,0.7){\vector(1,0){0.1}}
\put(1.7,0.2){\vector(1,0){0.1}}
\put(1.7,1.2){\vector(1,0){0.1}}
\linethickness{0.50mm}
\put(0,0.7){\line(1,0){0.5}}
\put(1,1.2){\line(1,0){1}}
\put(1,0.2){\line(1,0){1}}
\thinlines
\put(0.5,1.2){\oval(1,1)[br]}
\put(1,0.7){\circle{1}}
\put(0.25,0.9){\makebox(0,0)[b]{$#1$}}
\put(2.05,1.2){\makebox(0,0)[l]{$#2$}}
\put(2.05,0.2){\makebox(0,0)[l]{$#3$}}
\end{picture}
}}
\hfill}

\newcommand{\triangleThreexdot}[3]{
\mbox{\parbox{3cm}{\hspace{-0.1cm}
\begin{picture}(2.5,1.4)
\put(0.3,0.7){\vector(1,0){0.1}}
\put(1.7,0.2){\vector(1,0){0.1}}
\put(1.7,1.2){\vector(1,0){0.1}}
\linethickness{0.50mm}
\put(0,0.7){\line(1,0){0.5}}
\put(1,1.2){\line(1,0){1}}
\put(1,0.2){\line(1,0){1}}
\thinlines
\put(0.5,1.2){\oval(1,1)[br]}
\put(0.85,0.85){\circle*{0.15}}
\put(1,0.7){\circle{1}}
\put(0.25,0.9){\makebox(0,0)[b]{$#1$}}
\put(2.05,1.2){\makebox(0,0)[l]{$#2$}}
\put(2.05,0.2){\makebox(0,0)[l]{$#3$}}
\end{picture}
}}
\hfill}

\newcommand{\trianglebTwo}[3]{
\mbox{\parbox{3cm}{\hspace{-0.1cm}
\begin{picture}(2.5,1.4)
\put(0.3,0.7){\vector(-1,0){0.1}}
\put(1.9,0.2){\vector(-1,0){0.1}}
\put(1.9,1.2){\vector(-1,0){0.1}}
\put(0.5,0.7){\line(2,1){1}}
\put(0.5,0.7){\line(2,-1){1}}
\put(1.5,1.2){\line(0,-1){1}}
\put(1.5,1.2){\line(1,0){0.5}}
\linethickness{0.50mm}
\put(0,0.7){\line(1,0){0.5}}
\put(1.5,0.2){\line(1,0){0.5}}
\thinlines
\put(1.1,0.4){\line(1,2){0.4}}
\put(0.25,0.9){\makebox(0,0)[b]{$#1$}}
\put(2.05,1.2){\makebox(0,0)[l]{$#2$}}
\put(2.05,0.2){\makebox(0,0)[l]{$#3$}}
\end{picture}
}}
\hfill}

\newcommand{\trianglebThree}[3]{
\mbox{\parbox{3cm}{\hspace{-0.1cm}
\begin{picture}(2.5,1.4)
\put(0.3,0.7){\vector(-1,0){0.1}}
\put(1.9,0.2){\vector(1,0){0.1}}
\put(1.9,1.2){\vector(-1,0){0.1}}
\put(0.5,0.7){\line(2,1){1}}
\put(0.5,0.7){\line(2,-1){1}}
\put(1.5,1.2){\line(0,-1){1}}
\linethickness{0.50mm}
\put(0,0.7){\line(1,0){0.5}}
\put(1.5,1.2){\line(1,0){0.5}}
\put(1.5,0.2){\line(1,0){0.5}}
\thinlines
\put(1.5,0.2){\line(-1,2){0.4}}
\put(0.25,0.9){\makebox(0,0)[b]{$#1$}}
\put(2.05,1.2){\makebox(0,0)[l]{$#2$}}
\put(2.05,0.2){\makebox(0,0)[l]{$#3$}}
\end{picture}
}}
\hfill}

\newcommand{\trianglebThreex}[3]{
\mbox{\parbox{3cm}{\hspace{-0.1cm}
\begin{picture}(2.5,1.4)
\put(0.3,0.7){\vector(-1,0){0.1}}
\put(1.9,0.2){\vector(1,0){0.1}}
\put(1.9,1.2){\vector(-1,0){0.1}}
\put(0.5,0.7){\line(2,1){1}}
\put(0.5,0.7){\line(2,-1){1}}
\put(1.5,1.2){\line(0,-1){1}}
\linethickness{0.50mm}
\put(0,0.7){\line(1,0){0.5}}
\put(1.5,1.2){\line(1,0){0.5}}
\put(1.5,0.2){\line(1,0){0.5}}
\thinlines
\put(1.1,0.4){\line(1,2){0.4}}
\put(0.25,0.9){\makebox(0,0)[b]{$#1$}}
\put(2.05,1.2){\makebox(0,0)[l]{$#2$}}
\put(2.05,0.2){\makebox(0,0)[l]{$#3$}}
\end{picture}
}}
\hfill}

\newcommand{\boxbubblea}[4]{
\mbox{\parbox{3.5cm}{\hspace{-0.1cm}
\begin{picture}(3,1.4)
\put(1.5,0.7){\circle*{0.15}}
\put(0.8,0.2){\vector(1,0){0.1}}
\put(2.3,0.2){\vector(1,0){0.1}}
\put(2.3,1.2){\vector(1,0){0.1}}
\put(0.8,1.2){\vector(1,0){0.1}}
\put(1.5,1.2){\oval(1,1)[b]}
\put(0.5,0.2){\line(1,0){1.5}}
\put(0.5,1.2){\line(1,0){1.5}}
\linethickness{0.50mm}
\put(2.0,0.2){\line(1,0){0.5}}
\put(2.0,1.2){\line(1,0){0.5}}
\thinlines
\put(1,0.2){\line(0,1){1}}
\put(2,0.2){\line(0,1){1}}
\put(0.45,1.2){\makebox(0,0)[r]{$#1$}}
\put(0.45,0.2){\makebox(0,0)[r]{$#2$}}
\put(2.55,1.2){\makebox(0,0)[l]{$#3$}}
\put(2.55,0.2){\makebox(0,0)[l]{$#4$}}
\end{picture}
}} 
\hfill}

\newcommand{\boxbubbleb}[4]{
\mbox{\parbox{3.5cm}{\hspace{-0.1cm}
\begin{picture}(3,1.4)
\put(1.5,0.7){\circle*{0.15}}
\put(0.8,0.2){\vector(1,0){0.1}}
\put(2.3,0.2){\vector(1,0){0.1}}
\put(2.3,1.2){\vector(1,0){0.1}}
\put(0.8,1.2){\vector(1,0){0.1}}
\put(2.0,0.7){\oval(1,1)[l]}
\put(0.5,0.2){\line(1,0){1.5}}
\put(0.5,1.2){\line(1,0){1.5}}
\linethickness{0.50mm}
\put(2.0,0.2){\line(1,0){0.5}}
\put(2.0,1.2){\line(1,0){0.5}}
\thinlines
\put(1,0.2){\line(0,1){1}}
\put(2,0.2){\line(0,1){1}}
\put(0.45,1.2){\makebox(0,0)[r]{$#1$}}
\put(0.45,0.2){\makebox(0,0)[r]{$#2$}}
\put(2.55,1.2){\makebox(0,0)[l]{$#3$}}
\put(2.55,0.2){\makebox(0,0)[l]{$#4$}}
\end{picture}
}} 
\hfill}

\newcommand{\boxbubblec}[4]{
\mbox{\parbox{3.5cm}{\hspace{-0.1cm}
\begin{picture}(3,1.4)
\put(0.8,0.2){\vector(1,0){0.1}}
\put(2.3,0.2){\vector(1,0){0.1}}
\put(2.3,1.2){\vector(1,0){0.1}}
\put(0.8,1.2){\vector(1,0){0.1}}
\put(1.0,0.7){\oval(1,1)[r]}
\put(0.5,0.2){\line(1,0){1.5}}
\put(0.5,1.2){\line(1,0){1.5}}
\linethickness{0.50mm}
\put(2.0,0.2){\line(1,0){0.5}}
\put(2.0,1.2){\line(1,0){0.5}}
\thinlines
\put(1,0.2){\line(0,1){1}}
\put(2,0.2){\line(0,1){1}}
\put(0.45,1.2){\makebox(0,0)[r]{$#1$}}
\put(0.45,0.2){\makebox(0,0)[r]{$#2$}}
\put(2.55,1.2){\makebox(0,0)[l]{$#3$}}
\put(2.55,0.2){\makebox(0,0)[l]{$#4$}}
\end{picture}
}} 
\hfill}

\newcommand{\boxbubblecdot}[4]{
\mbox{\parbox{3.5cm}{\hspace{-0.1cm}
\begin{picture}(3,1.4)
\put(0.8,0.2){\vector(1,0){0.1}}
\put(2.3,0.2){\vector(1,0){0.1}}
\put(2.3,1.2){\vector(1,0){0.1}}
\put(0.8,1.2){\vector(1,0){0.1}}
\put(1.0,0.7){\oval(1,1)[r]}
\put(0.5,0.2){\line(1,0){1.5}}
\put(0.5,1.2){\line(1,0){1.5}}
\linethickness{0.50mm}
\put(2.0,0.2){\line(1,0){0.5}}
\put(2.0,1.2){\line(1,0){0.5}}
\thinlines
\put(1,0.2){\line(0,1){1}}
\put(2,0.2){\line(0,1){1}}
\put(1.5,0.7){\circle*{0.15}}
\put(0.45,1.2){\makebox(0,0)[r]{$#1$}}
\put(0.45,0.2){\makebox(0,0)[r]{$#2$}}
\put(2.55,1.2){\makebox(0,0)[l]{$#3$}}
\put(2.55,0.2){\makebox(0,0)[l]{$#4$}}
\end{picture}
}} 
\hfill}

\newcommand{\boxxbNLO}[4]{
\mbox{\parbox{3.5cm}{\hspace{-0.1cm}
\begin{picture}(3,1.4)
\put(0.8,0.2){\vector(1,0){0.1}}
\put(2.3,0.2){\vector(1,0){0.1}}
\put(2.3,1.2){\vector(1,0){0.1}}
\put(0.8,1.2){\vector(1,0){0.1}}
\put(1,0.2){\line(1,1){1}}
\put(0.5,0.2){\line(1,0){1.5}}
\put(0.5,1.2){\line(1,0){1.5}}
\linethickness{0.50mm}
\put(2.0,0.2){\line(1,0){0.5}}
\put(2.0,1.2){\line(1,0){0.5}}
\thinlines
\put(1,0.2){\line(0,1){1}}
\put(2,0.2){\line(0,1){1}}
\put(0.45,1.2){\makebox(0,0)[r]{$#1$}}
\put(0.45,0.2){\makebox(0,0)[r]{$#2$}}
\put(2.55,1.2){\makebox(0,0)[l]{$#3$}}
\put(2.55,0.2){\makebox(0,0)[l]{$#4$}}
\end{picture}
}} 
\hfill}

\newcommand{\boxxbdotNLO}[4]{
\mbox{\parbox{3.5cm}{\hspace{-0.1cm}
\begin{picture}(3,1.4)
\put(0.8,0.2){\vector(1,0){0.1}}
\put(2.3,0.2){\vector(1,0){0.1}}
\put(2.3,1.2){\vector(1,0){0.1}}
\put(0.8,1.2){\vector(1,0){0.1}}
\put(1,0.2){\line(1,1){1}}
\put(0.5,0.2){\line(1,0){1.5}}
\put(0.5,1.2){\line(1,0){1.5}}
\linethickness{0.50mm}
\put(2.0,0.2){\line(1,0){0.5}}
\put(2.0,1.2){\line(1,0){0.5}}
\thinlines
\put(2.0,0.7){\circle*{0.15}}
\put(1,0.2){\line(0,1){1}}
\put(2,0.2){\line(0,1){1}}
\put(0.45,1.2){\makebox(0,0)[r]{$#1$}}
\put(0.45,0.2){\makebox(0,0)[r]{$#2$}}
\put(2.55,1.2){\makebox(0,0)[l]{$#3$}}
\put(2.55,0.2){\makebox(0,0)[l]{$#4$}}
\end{picture}
}} 
\hfill}

\newcommand{\boxxbpaNLO}[4]{
\mbox{\parbox{3.5cm}{\hspace{-0.1cm}
\begin{picture}(3,1.4)
\put(0.8,0.2){\vector(1,0){0.1}}
\put(2.3,0.2){\vector(1,0){0.1}}
\put(2.3,1.2){\vector(1,0){0.1}}
\put(0.8,1.2){\vector(1,0){0.1}}
\put(0.5,0.2){\line(1,0){1.5}}
\put(0.5,1.2){\line(1,0){1.5}}
\linethickness{0.50mm}
\put(2.0,0.2){\line(1,0){0.5}}
\put(2.0,1.2){\line(1,0){0.5}}
\thinlines
\put(1.5,1.2){\line(1,-2){0.5}}
\put(1,0.2){\line(0,1){1}}
\put(2,0.2){\line(0,1){1}}
\put(0.45,1.2){\makebox(0,0)[r]{$#1$}}
\put(0.45,0.2){\makebox(0,0)[r]{$#2$}}
\put(2.55,1.2){\makebox(0,0)[l]{$#3$}}
\put(2.55,0.2){\makebox(0,0)[l]{$#4$}}
\end{picture}
}} 
\hfill}

\newcommand{\boxxbpbNLO}[4]{
\mbox{\parbox{3.5cm}{\hspace{-0.1cm}
\begin{picture}(3,1.4)
\put(0.8,0.2){\vector(1,0){0.1}}
\put(2.3,0.2){\vector(1,0){0.1}}
\put(2.3,1.2){\vector(1,0){0.1}}
\put(0.8,1.2){\vector(1,0){0.1}}
\put(0.5,0.2){\line(1,0){1.5}}
\put(0.5,1.2){\line(1,0){1.5}}
\linethickness{0.50mm}
\put(2.0,0.2){\line(1,0){0.5}}
\put(2.0,1.2){\line(1,0){0.5}}
\thinlines
\put(1.0,1.2){\line(2,-1){1}}
\put(1,0.2){\line(0,1){1}}
\put(2,0.2){\line(0,1){1}}
\put(0.45,1.2){\makebox(0,0)[r]{$#1$}}
\put(0.45,0.2){\makebox(0,0)[r]{$#2$}}
\put(2.55,1.2){\makebox(0,0)[l]{$#3$}}
\put(2.55,0.2){\makebox(0,0)[l]{$#4$}}
\end{picture}
}} 
\hfill}

\newcommand{\doubleboxa}[4]{
\mbox{\parbox{3.5cm}{\hspace{-0.1cm}
\begin{picture}(3,1.4)
\put(0.8,0.2){\vector(1,0){0.1}}
\put(2.3,0.2){\vector(1,0){0.1}}
\put(2.3,1.2){\vector(1,0){0.1}}
\put(0.8,1.2){\vector(1,0){0.1}}
\put(0.5,0.2){\line(1,0){1.5}}
\put(0.5,1.2){\line(1,0){1.5}}
\linethickness{0.50mm}
\put(2.0,0.2){\line(1,0){0.5}}
\put(2.0,1.2){\line(1,0){0.5}}
\thinlines
\put(1,0.2){\line(0,1){1}}
\put(1.5,0.2){\line(0,1){1}}
\put(2,0.2){\line(0,1){1}}
\put(0.45,1.2){\makebox(0,0)[r]{$#1$}}
\put(0.45,0.2){\makebox(0,0)[r]{$#2$}}
\put(2.55,1.2){\makebox(0,0)[l]{$#3$}}
\put(2.55,0.2){\makebox(0,0)[l]{$#4$}}
\end{picture}
}}
\hfill}

\newcommand{\doubleboxatwo}[4]{
\mbox{\parbox{3.5cm}{\hspace{-0.1cm}
\begin{picture}(3,1.4)
\put(0.8,0.2){\vector(1,0){0.1}}
\put(2.3,0.2){\vector(1,0){0.1}}
\put(2.3,1.2){\vector(1,0){0.1}}
\put(0.8,1.2){\vector(1,0){0.1}}
\put(0.5,0.2){\line(1,0){1.5}}
\put(0.5,1.2){\line(1,0){1.5}}
\linethickness{0.50mm}
\put(2.0,0.2){\line(1,0){0.5}}
\put(2.0,1.2){\line(1,0){0.5}}
\thinlines
\put(1,0.2){\line(0,1){1}}
\put(1.5,0.2){\line(0,1){1}}
\put(2,0.2){\line(0,1){1}}
\put(0.45,1.2){\makebox(0,0)[r]{$#1$}}
\put(0.45,0.2){\makebox(0,0)[r]{$#2$}}
\put(2.55,1.2){\makebox(0,0)[l]{$#3$}}
\put(2.55,0.2){\makebox(0,0)[l]{$#4$}}
\put(2.085,0.7){\makebox(0,0)[l]{${_{(l-p_{123})^2}}$}}
\end{picture}
}}
\hfill}

\newcommand{\doubleboxathree}[4]{
\mbox{\parbox{3.5cm}{\hspace{-0.1cm}
\begin{picture}(3,1.4)
\put(0.8,0.2){\vector(1,0){0.1}}
\put(2.3,0.2){\vector(1,0){0.1}}
\put(2.3,1.2){\vector(1,0){0.1}}
\put(0.8,1.2){\vector(1,0){0.1}}
\put(0.5,0.2){\line(1,0){1.5}}
\put(0.5,1.2){\line(1,0){1.5}}
\linethickness{0.50mm}
\put(2.0,0.2){\line(1,0){0.5}}
\put(2.0,1.2){\line(1,0){0.5}}
\thinlines
\put(1,0.2){\line(0,1){1}}
\put(1.5,0.2){\line(0,1){1}}
\put(2,0.2){\line(0,1){1}}
\put(0.45,1.2){\makebox(0,0)[r]{$#1$}}
\put(0.45,0.2){\makebox(0,0)[r]{$#2$}}
\put(2.55,1.2){\makebox(0,0)[l]{$#3$}}
\put(2.55,0.2){\makebox(0,0)[l]{$#4$}}
\put(2.085,0.7){\makebox(0,0)[l]{${_{(k-p_{1})^2}}$}}
\end{picture}
}}
\hfill}

\newcommand{\doubleboxb}[4]{
\mbox{\parbox{3.5cm}{\hspace{-0.1cm}
\begin{picture}(3,1.4)
\put(0.8,0.2){\vector(1,0){0.1}}
\put(2.3,0.2){\vector(1,0){0.1}}
\put(2.3,1.2){\vector(1,0){0.1}}
\put(0.8,1.2){\vector(1,0){0.1}}
\put(0.5,0.2){\line(1,0){1.5}}
\put(0.5,1.2){\line(1,0){1.5}}
\linethickness{0.50mm}
\put(2.0,0.2){\line(1,0){0.5}}
\put(2.0,1.2){\line(1,0){0.5}}
\thinlines
\put(1,0.2){\line(0,1){1}}
\put(1.0,0.7){\line(1,0){1}}
\put(2,0.2){\line(0,1){1}}
\put(0.45,1.2){\makebox(0,0)[r]{$#1$}}
\put(0.45,0.2){\makebox(0,0)[r]{$#2$}}
\put(2.55,1.2){\makebox(0,0)[l]{$#3$}}
\put(2.55,0.2){\makebox(0,0)[l]{$#4$}}
\end{picture}
}}
\hfill}

\newcommand{\doubleboxbtwo}[4]{
\mbox{\parbox{3.5cm}{\hspace{-0.1cm}
\begin{picture}(3,1.4)
\put(0.8,0.2){\vector(1,0){0.1}}
\put(2.3,0.2){\vector(1,0){0.1}}
\put(2.3,1.2){\vector(1,0){0.1}}
\put(0.8,1.2){\vector(1,0){0.1}}
\put(0.5,0.2){\line(1,0){1.5}}
\put(0.5,1.2){\line(1,0){1.5}}
\linethickness{0.50mm}
\put(2.0,0.2){\line(1,0){0.5}}
\put(2.0,1.2){\line(1,0){0.5}}
\thinlines
\put(1,0.2){\line(0,1){1}}
\put(1.0,0.7){\line(1,0){1}}
\put(2,0.2){\line(0,1){1}}
\put(0.45,1.2){\makebox(0,0)[r]{$#1$}}
\put(0.45,0.2){\makebox(0,0)[r]{$#2$}}
\put(2.55,1.2){\makebox(0,0)[l]{$#3$}}
\put(2.55,0.2){\makebox(0,0)[l]{$#4$}}
\put(2.085,0.7){\makebox(0,0)[l]{${_{(k)^2}}$}}
\end{picture}
}}
\hfill}

\newcommand{\boxbubbleOB}[4]{
\mbox{\parbox{3.5cm}{\hspace{0.0cm}
\begin{picture}(3.5,1.4)
\put(0.8,0.2){\vector(-1,0){0.1}}
\put(2.3,0.2){\vector(-1,0){0.1}}
\put(2.3,1.2){\vector(1,0){0.1}}
\put(0.8,1.2){\vector(1,0){0.1}}
\put(0.75,0.45){\circle*{0.15}}
\put(1,0.45){\circle{0.5}}
\put(0.5,0.2){\line(1,0){2.0}}
\put(0.5,1.2){\line(1,0){2.0}}
\put(1,0.7){\line(0,1){0.5}}
\put(2,1.2){\line(0,-1){1.0}}

\linethickness{0.50mm}
\put(0.5,0.2){\line(1,0){0.5}}
\put(2.0,1.2){\line(1,0){0.5}}
\thinlines
\put(0.45,1.2){\makebox(0,0)[r]{$#1$}}
\put(0.45,0.2){\makebox(0,0)[r]{$#2$}}
\put(2.55,1.2){\makebox(0,0)[l]{$#3$}}
\put(2.55,0.2){\makebox(0,0)[l]{$#4$}}

\end{picture}
}} 
\hfill}

\newcommand{\boxbubbleaB}[4]{
\mbox{\parbox{3.5cm}{\hspace{-0.1cm}
\begin{picture}(3,1.4)
\put(1.5,0.7){\circle*{0.15}}
\put(0.8,0.2){\vector(-1,0){0.1}}
\put(2.3,0.2){\vector(-1,0){0.1}}
\put(2.3,1.2){\vector(1,0){0.1}}
\put(0.8,1.2){\vector(1,0){0.1}}
\put(1.5,1.2){\oval(1,1)[b]}
\put(1.0,0.2){\line(1,0){1.5}}
\put(0.5,1.2){\line(1,0){1.5}}
\linethickness{0.50mm}
\put(0.5,0.2){\line(1,0){0.5}}
\put(2.0,1.2){\line(1,0){0.5}}
\thinlines
\put(1,0.2){\line(0,1){1}}
\put(2,0.2){\line(0,1){1}}
\put(0.45,1.2){\makebox(0,0)[r]{$#1$}}
\put(0.45,0.2){\makebox(0,0)[r]{$#2$}}
\put(2.55,1.2){\makebox(0,0)[l]{$#3$}}
\put(2.55,0.2){\makebox(0,0)[l]{$#4$}}
\end{picture}
}} 
\hfill}

\newcommand{\boxxaB}[4]{
\mbox{\parbox{3.5cm}{\hspace{-0.1cm}
\begin{picture}(3,1.4)
\put(0.8,0.2){\vector(-1,0){0.1}}
\put(2.3,0.2){\vector(-1,0){0.1}}
\put(2.3,1.2){\vector(1,0){0.1}}
\put(0.8,1.2){\vector(1,0){0.1}}
\put(1,0.2){\line(1,1){1}}
\put(1.0,0.2){\line(1,0){1.5}}
\put(0.5,1.2){\line(1,0){1.5}}
\linethickness{0.50mm}
\put(0.5,0.2){\line(1,0){0.5}}
\put(2.0,1.2){\line(1,0){0.5}}
\thinlines
\put(1,0.2){\line(0,1){1}}
\put(2,0.2){\line(0,1){1}}
\put(0.45,1.2){\makebox(0,0)[r]{$#1$}}
\put(0.45,0.2){\makebox(0,0)[r]{$#2$}}
\put(2.55,1.2){\makebox(0,0)[l]{$#3$}}
\put(2.55,0.2){\makebox(0,0)[l]{$#4$}}
\end{picture}
}} 
\hfill}

\newcommand{\boxxbB}[4]{
\mbox{\parbox{3.5cm}{\hspace{-0.1cm}
\begin{picture}(3,1.4)
\put(0.8,0.2){\vector(-1,0){0.1}}
\put(2.3,0.2){\vector(-1,0){0.1}}
\put(2.3,1.2){\vector(1,0){0.1}}
\put(0.8,1.2){\vector(1,0){0.1}}
\put(2,0.2){\line(-1,1){1}}
\put(1.0,0.2){\line(1,0){1.5}}
\put(0.5,1.2){\line(1,0){1.5}}
\linethickness{0.50mm}
\put(0.5,0.2){\line(1,0){0.5}}
\put(2.0,1.2){\line(1,0){0.5}}
\thinlines
\put(1,0.2){\line(0,1){1}}
\put(2,0.2){\line(0,1){1}}
\put(0.45,1.2){\makebox(0,0)[r]{$#1$}}
\put(0.45,0.2){\makebox(0,0)[r]{$#2$}}
\put(2.55,1.2){\makebox(0,0)[l]{$#3$}}
\put(2.55,0.2){\makebox(0,0)[l]{$#4$}}
\end{picture}
}} 
\hfill}

\newcommand{\boxxbbB}[4]{
\mbox{\parbox{3.5cm}{\hspace{-0.1cm}
\begin{picture}(3,1.4)
\put(0.8,0.2){\vector(-1,0){0.1}}
\put(2.3,0.2){\vector(-1,0){0.1}}
\put(2.3,1.2){\vector(1,0){0.1}}
\put(0.8,1.2){\vector(1,0){0.1}}
\put(1.5,0.7){\circle*{0.15}}
\put(2,0.2){\line(-1,1){1}}
\put(1.0,0.2){\line(1,0){1.5}}
\put(0.5,1.2){\line(1,0){1.5}}
\linethickness{0.50mm}
\put(0.5,0.2){\line(1,0){0.5}}
\put(2.0,1.2){\line(1,0){0.5}}
\thinlines
\put(1,0.2){\line(0,1){1}}
\put(2,0.2){\line(0,1){1}}
\put(0.45,1.2){\makebox(0,0)[r]{$#1$}}
\put(0.45,0.2){\makebox(0,0)[r]{$#2$}}
\put(2.55,1.2){\makebox(0,0)[l]{$#3$}}
\put(2.55,0.2){\makebox(0,0)[l]{$#4$}}
\end{picture}
}} 
\hfill}

\newcommand{\boxxbbbbbB}[4]{
\mbox{\parbox{3.5cm}{\hspace{-0.1cm}
\begin{picture}(3,1.4)
\put(0.8,0.2){\vector(-1,0){0.1}}
\put(2.3,0.2){\vector(-1,0){0.1}}
\put(2.3,1.2){\vector(1,0){0.1}}
\put(0.8,1.2){\vector(1,0){0.1}}
\put(1.5,1.2){\circle*{0.15}}
\put(2,0.2){\line(-1,1){1}}
\put(1.0,0.2){\line(1,0){1.5}}
\put(0.5,1.2){\line(1,0){1.5}}
\linethickness{0.50mm}
\put(0.5,0.2){\line(1,0){0.5}}
\put(2.0,1.2){\line(1,0){0.5}}
\thinlines
\put(1,0.2){\line(0,1){1}}
\put(2,0.2){\line(0,1){1}}
\put(0.45,1.2){\makebox(0,0)[r]{$#1$}}
\put(0.45,0.2){\makebox(0,0)[r]{$#2$}}
\put(2.55,1.2){\makebox(0,0)[l]{$#3$}}
\put(2.55,0.2){\makebox(0,0)[l]{$#4$}}
\end{picture}
}} 
\hfill}

\newcommand{\boxxbbbbbbB}[4]{
\mbox{\parbox{3.5cm}{\hspace{-0.1cm}
\begin{picture}(3,1.4)
\put(0.8,0.2){\vector(-1,0){0.1}}
\put(2.3,0.2){\vector(-1,0){0.1}}
\put(2.3,1.2){\vector(1,0){0.1}}
\put(0.8,1.2){\vector(1,0){0.1}}
\put(1.0,0.7){\circle*{0.15}}
\put(2,0.2){\line(-1,1){1}}
\put(1.0,0.2){\line(1,0){1.5}}
\put(0.5,1.2){\line(1,0){1.5}}
\linethickness{0.50mm}
\put(0.5,0.2){\line(1,0){0.5}}
\put(2.0,1.2){\line(1,0){0.5}}
\thinlines
\put(1,0.2){\line(0,1){1}}
\put(2,0.2){\line(0,1){1}}
\put(0.45,1.2){\makebox(0,0)[r]{$#1$}}
\put(0.45,0.2){\makebox(0,0)[r]{$#2$}}
\put(2.55,1.2){\makebox(0,0)[l]{$#3$}}
\put(2.55,0.2){\makebox(0,0)[l]{$#4$}}
\end{picture}
}} 
\hfill}

\newcommand{\boxxbpaB}[4]{
\mbox{\parbox{3.5cm}{\hspace{-0.1cm}
\begin{picture}(3,1.4)
\put(0.8,0.2){\vector(-1,0){0.1}}
\put(2.3,0.2){\vector(-1,0){0.1}}
\put(2.3,1.2){\vector(1,0){0.1}}
\put(0.8,1.2){\vector(1,0){0.1}}
\put(1.0,0.2){\line(1,0){1.5}}
\put(0.5,1.2){\line(1,0){1.5}}
\linethickness{0.50mm}
\put(0.5,0.2){\line(1,0){0.5}}
\put(2.0,1.2){\line(1,0){0.5}}
\thinlines
\put(1.5,1.2){\line(1,-2){0.5}}
\put(1,0.2){\line(0,1){1}}
\put(2,0.2){\line(0,1){1}}
\put(0.45,1.2){\makebox(0,0)[r]{$#1$}}
\put(0.45,0.2){\makebox(0,0)[r]{$#2$}}
\put(2.55,1.2){\makebox(0,0)[l]{$#3$}}
\put(2.55,0.2){\makebox(0,0)[l]{$#4$}}
\end{picture}
}} 
\hfill}

\newcommand{\doubleboxB}[4]{
\mbox{\parbox{3.5cm}{\hspace{-0.1cm}
\begin{picture}(3,1.4)
\put(0.8,0.2){\vector(-1,0){0.1}}
\put(2.3,0.2){\vector(-1,0){0.1}}
\put(2.3,1.2){\vector(1,0){0.1}}
\put(0.8,1.2){\vector(1,0){0.1}}
\put(1.0,0.2){\line(1,0){1.5}}
\put(0.5,1.2){\line(1,0){1.5}}
\linethickness{0.50mm}
\put(0.5,0.2){\line(1,0){0.5}}
\put(2.0,1.2){\line(1,0){0.5}}
\thinlines
\put(1,0.2){\line(0,1){1}}
\put(1.5,0.2){\line(0,1){1}}
\put(2,0.2){\line(0,1){1}}
\put(0.45,1.2){\makebox(0,0)[r]{$#1$}}
\put(0.45,0.2){\makebox(0,0)[r]{$#2$}}
\put(2.55,1.2){\makebox(0,0)[l]{$#3$}}
\put(2.55,0.2){\makebox(0,0)[l]{$#4$}}
\end{picture}
}}
\hfill}

\newcommand{\doubleboxBtwo}[4]{
\mbox{\parbox{3.5cm}{\hspace{-0.1cm}
\begin{picture}(3,1.4)
\put(0.8,0.2){\vector(-1,0){0.1}}
\put(2.3,0.2){\vector(-1,0){0.1}}
\put(2.3,1.2){\vector(1,0){0.1}}
\put(0.8,1.2){\vector(1,0){0.1}}
\put(1.0,0.2){\line(1,0){1.5}}
\put(0.5,1.2){\line(1,0){1.5}}
\linethickness{0.50mm}
\put(0.5,0.2){\line(1,0){0.5}}
\put(2.0,1.2){\line(1,0){0.5}}
\thinlines
\put(1,0.2){\line(0,1){1}}
\put(1.5,0.2){\line(0,1){1}}
\put(2,0.2){\line(0,1){1}}
\put(0.45,1.2){\makebox(0,0)[r]{$#1$}}
\put(0.45,0.2){\makebox(0,0)[r]{$#2$}}
\put(2.55,1.2){\makebox(0,0)[l]{$#3$}}
\put(2.55,0.2){\makebox(0,0)[l]{$#4$}}
\put(2.085,0.7){\makebox(0,0)[l]{${_{(k-p_1)^2}}$}}
\end{picture}
}}
\hfill}

\newcommand{\trianglecrossA}[3]{
\mbox{\parbox{3cm}{\hspace{-0.1cm}
\begin{picture}(2.5,1.4)
\put(0.3,0.7){\vector(1,0){0.1}}
\put(1.9,0.2){\vector(1,0){0.1}}
\put(1.9,1.2){\vector(1,0){0.1}}
\put(0.5,0.7){\line(1,1){0.5}}
\put(0.5,0.7){\line(1,-1){0.5}}
\put(1.5,1.2){\line(-1,-2){0.5}}
\put(1.0,1.2){\line(1,-2){0.18}}
\put(1.5,0.2){\line(-1,2){0.18}}
\put(1,1.2){\line(1,0){0.5}}
\put(1,0.2){\line(1,0){0.5}}
\linethickness{0.50mm}
\put(0,0.7){\line(1,0){0.5}}
\put(1.5,1.2){\line(1,0){0.5}}
\put(1.5,0.2){\line(1,0){0.5}}
\thinlines
\put(0.25,0.9){\makebox(0,0)[b]{$#1$}}
\put(2.05,1.2){\makebox(0,0)[l]{$#2$}}
\put(2.05,0.2){\makebox(0,0)[l]{$#3$}}
\end{picture}
}}
\hfill}

\newcommand{\trianglecrossB}[3]{
\mbox{\parbox{3cm}{\hspace{-0.1cm}
\begin{picture}(2.5,1.4)
\put(0.3,0.7){\vector(1,0){0.1}}
\put(1.9,0.2){\vector(1,0){0.1}}
\put(1.9,1.2){\vector(1,0){0.1}}
\put(0.5,0.7){\line(1,1){0.5}}
\put(0.5,0.7){\line(1,-1){0.5}}
\put(1.5,1.2){\line(-1,-2){0.5}}
\put(1.0,1.2){\line(1,-2){0.18}}
\put(1.5,0.2){\line(-1,2){0.18}}
\put(1,1.2){\line(1,0){0.5}}
\put(1,0.2){\line(1,0){0.5}}
\put(0,0.7){\line(1,0){0.5}}
\linethickness{0.50mm}
\put(1.5,1.2){\line(1,0){0.5}}
\put(1.5,0.2){\line(1,0){0.5}}
\thinlines
\put(0.25,0.9){\makebox(0,0)[b]{$#1$}}
\put(2.05,1.2){\makebox(0,0)[l]{$#2$}}
\put(2.05,0.2){\makebox(0,0)[l]{$#3$}}
\end{picture}
}}
\hfill}

\newcommand{\trianglecrossC}[3]{
\mbox{\parbox{3cm}{\hspace{-0.1cm}
\begin{picture}(2.5,1.4)
\put(0.3,0.7){\vector(1,0){0.1}}
\put(1.9,0.2){\vector(1,0){0.1}}
\put(1.9,1.2){\vector(1,0){0.1}}
\put(0.5,0.7){\line(1,1){0.5}}
\put(0.5,0.7){\line(1,-1){0.5}}
\put(1.5,1.2){\line(-1,-2){0.5}}
\put(1.0,1.2){\line(1,-2){0.18}}
\put(1.5,0.2){\line(-1,2){0.18}}
\put(1,1.2){\line(1,0){0.5}}
\put(1,0.2){\line(1,0){0.5}}
\linethickness{0.50mm}
\put(0,0.7){\line(1,0){0.5}}
\thinlines
\put(1.5,1.2){\line(1,0){0.5}}
\put(1.5,0.2){\line(1,0){0.5}}
\put(0.25,0.9){\makebox(0,0)[b]{$#1$}}
\put(2.05,1.2){\makebox(0,0)[l]{$#2$}}
\put(2.05,0.2){\makebox(0,0)[l]{$#3$}}
\end{picture}
}}
\hfill}

\newcommand{\boxxbmcrossxAa}[4]{
\mbox{\parbox{3.5cm}{\hspace{-0.1cm}
\begin{picture}(3,1.4)
\put(0.8,0.2){\vector(-1,0){0.1}}
\put(2.3,0.2){\vector(1,0){0.1}}
\put(2.3,1.2){\vector(-1,0){0.1}}
\put(0.8,1.2){\vector(1,0){0.1}}
\linethickness{0.50mm}
\put(0.5,0.2){\line(1,0){1}}
\put(2.0,0.2){\line(1,0){0.5}}
\thinlines
\put(1.5,0.2){\line(1,0){0.5}}
\put(1.5,0.2){\line(1,2){0.2}}
\put(2,1.2){\line(-1,-2){0.2}}
\put(0.5,1.2){\line(1,0){2}}
\put(1,1.2){\line(1,-2){0.5}}
\put(1.5,1.2){\line(1,-2){0.5}}
\put(0.45,1.2){\makebox(0,0)[r]{$#1$}}
\put(0.45,0.2){\makebox(0,0)[r]{$#2$}}
\put(2.55,1.2){\makebox(0,0)[l]{$#3$}}
\put(2.55,0.2){\makebox(0,0)[l]{$#4$}}
\end{picture}
}} 
\hfill}

\newcommand{\boxxbmcrossxAb}[4]{
\mbox{\parbox{3.5cm}{\hspace{-0.1cm}
\begin{picture}(3,1.4)
\put(0.8,0.2){\vector(-1,0){0.1}}
\put(2.3,0.2){\vector(1,0){0.1}}
\put(2.3,1.2){\vector(-1,0){0.1}}
\put(0.8,1.2){\vector(1,0){0.1}}
\linethickness{0.50mm}
\put(0.5,0.2){\line(1,0){1}}
\put(2.0,0.2){\line(1,0){0.5}}
\thinlines
\put(1.5,0.2){\line(1,0){0.5}}
\put(1.5,0.2){\line(1,2){0.2}}
\put(2,1.2){\line(-1,-2){0.2}}
\put(0.5,1.2){\line(1,0){2}}
\put(1,1.2){\line(1,-2){0.5}}
\put(1.5,1.2){\line(1,-2){0.5}}
\put(0.45,1.2){\makebox(0,0)[r]{$#1$}}
\put(0.45,0.2){\makebox(0,0)[r]{$#2$}}
\put(2.55,1.2){\makebox(0,0)[l]{$#3$}}
\put(2.55,0.2){\makebox(0,0)[l]{$#4$}}
\put(2.085,0.7){\makebox(0,0)[l]{${_{(k)^2}}$}}
\end{picture}
}} 
\hfill}

\newcommand{\boxxbmcrossxBa}[4]{
\mbox{\parbox{3.5cm}{\hspace{-0.1cm}
\begin{picture}(3,1.4)
\put(0.8,0.2){\vector(1,0){0.1}}
\put(2.3,0.2){\vector(1,0){0.1}}
\put(2.3,1.2){\vector(1,0){0.1}}
\put(0.8,1.2){\vector(1,0){0.1}}
\put(0.5,0.2){\line(1,0){1}}
\linethickness{0.50mm}
\put(2.0,1.2){\line(1,0){0.5}}
\put(2.0,0.2){\line(1,0){0.5}}
\thinlines
\put(2,1.2){\line(-1,-2){0.2}}
\put(1.5,0.2){\line(1,0){0.5}}
\put(1.5,0.2){\line(1,2){0.2}}
\put(0.5,1.2){\line(1,0){1.5}}
\put(1,1.2){\line(1,-2){0.5}}
\put(1.5,1.2){\line(1,-2){0.5}}
\put(0.45,1.2){\makebox(0,0)[r]{$#1$}}
\put(0.45,0.2){\makebox(0,0)[r]{$#2$}}
\put(2.55,1.2){\makebox(0,0)[l]{$#3$}}
\put(2.55,0.2){\makebox(0,0)[l]{$#4$}}
\end{picture}
}} 
\hfill}

\newcommand{\boxxbmcrossxBb}[4]{
\mbox{\parbox{3.5cm}{\hspace{-0.1cm}
\begin{picture}(3,1.4)
\put(0.8,0.2){\vector(1,0){0.1}}
\put(2.3,0.2){\vector(1,0){0.1}}
\put(2.3,1.2){\vector(1,0){0.1}}
\put(0.8,1.2){\vector(1,0){0.1}}
\put(0.5,0.2){\line(1,0){1}}
\linethickness{0.50mm}
\put(2.0,1.2){\line(1,0){0.5}}
\put(2.0,0.2){\line(1,0){0.5}}
\thinlines
\put(2,1.2){\line(-1,-2){0.2}}
\put(1.5,0.2){\line(1,0){0.5}}
\put(1.5,0.2){\line(1,2){0.2}}
\put(0.5,1.2){\line(1,0){1.5}}
\put(1,1.2){\line(1,-2){0.5}}
\put(1.5,1.2){\line(1,-2){0.5}}
\put(0.45,1.2){\makebox(0,0)[r]{$#1$}}
\put(0.45,0.2){\makebox(0,0)[r]{$#2$}}
\put(2.55,1.2){\makebox(0,0)[l]{$#3$}}
\put(2.55,0.2){\makebox(0,0)[l]{$#4$}}
\put(2.085,0.7){\makebox(0,0)[l]{${_{(k-l-p_{12})^2}}$}}
\end{picture}
}} 
\hfill}

\newcommand{\boxxbmcrossxBc}[4]{
\mbox{\parbox{3.5cm}{\hspace{-0.1cm}
\begin{picture}(3,1.4)
\put(0.8,0.2){\vector(1,0){0.1}}
\put(2.3,0.2){\vector(1,0){0.1}}
\put(2.3,1.2){\vector(1,0){0.1}}
\put(0.8,1.2){\vector(1,0){0.1}}
\put(0.5,0.2){\line(1,0){1}}
\linethickness{0.50mm}
\put(2.0,1.2){\line(1,0){0.5}}
\put(2.0,0.2){\line(1,0){0.5}}
\thinlines
\put(2,1.2){\line(-1,-2){0.2}}
\put(1.5,0.2){\line(1,0){0.5}}
\put(1.5,0.2){\line(1,2){0.2}}
\put(0.5,1.2){\line(1,0){1.5}}
\put(1,1.2){\line(1,-2){0.5}}
\put(1.5,1.2){\line(1,-2){0.5}}
\put(0.45,1.2){\makebox(0,0)[r]{$#1$}}
\put(0.45,0.2){\makebox(0,0)[r]{$#2$}}
\put(2.55,1.2){\makebox(0,0)[l]{$#3$}}
\put(2.55,0.2){\makebox(0,0)[l]{$#4$}}
\put(2.085,0.7){\makebox(0,0)[l]{${_{(k-p_{12})^2}}$}}
\end{picture}
}} 
\hfill}

\newcommand{\boxxbmcrossxBd}[4]{
\mbox{\parbox{3.5cm}{\hspace{-0.1cm}
\begin{picture}(3,1.4)
\put(0.8,0.2){\vector(1,0){0.1}}
\put(2.3,0.2){\vector(1,0){0.1}}
\put(2.3,1.2){\vector(1,0){0.1}}
\put(0.8,1.2){\vector(1,0){0.1}}
\put(0.5,0.2){\line(1,0){1}}
\linethickness{0.50mm}
\put(2.0,1.2){\line(1,0){0.5}}
\put(2.0,0.2){\line(1,0){0.5}}
\thinlines
\put(2,1.2){\line(-1,-2){0.2}}
\put(1.5,0.2){\line(1,0){0.5}}
\put(1.5,0.2){\line(1,2){0.2}}
\put(0.5,1.2){\line(1,0){1.5}}
\put(1,1.2){\line(1,-2){0.5}}
\put(1.5,1.2){\line(1,-2){0.5}}
\put(0.45,1.2){\makebox(0,0)[r]{$#1$}}
\put(0.45,0.2){\makebox(0,0)[r]{$#2$}}
\put(2.55,1.2){\makebox(0,0)[l]{$#3$}}
\put(2.55,0.2){\makebox(0,0)[l]{$#4$}}
\put(2.085,0.7){\makebox(0,0)[l]{${_{(l-p_1)^2}}$}}
\end{picture}
}} 
\hfill}

\newcommand{\doublecrossA}[4]{
\mbox{\parbox{3.5cm}{\hspace{-0.1cm}
\begin{picture}(3,1.4)
\put(0.8,0.2){\vector(1,0){0.1}}
\put(2.3,0.2){\vector(1,0){0.1}}
\put(2.3,1.2){\vector(1,0){0.1}}
\put(0.8,1.2){\vector(1,0){0.1}}
\put(0.5,0.2){\line(1,0){2}}
\put(1.5,0.2){\line(1,2){0.2}}
\put(2,1.2){\line(-1,-2){0.2}}
\put(0.5,1.2){\line(1,0){2}}
\put(1,0.2){\line(0,1){1}}
\put(1.5,1.2){\line(1,-2){0.5}}
\linethickness{0.50mm}
\put(2.0,1.2){\line(1,0){0.5}}
\put(2.0,0.2){\line(1,0){0.5}}
\thinlines
\put(0.45,1.2){\makebox(0,0)[r]{$#1$}}
\put(0.45,0.2){\makebox(0,0)[r]{$#2$}}
\put(2.55,1.2){\makebox(0,0)[l]{$#3$}}
\put(2.55,0.2){\makebox(0,0)[l]{$#4$}}
\end{picture}
}} 
\hfill}

\newcommand{\doublecrossAb}[4]{
\mbox{\parbox{3.5cm}{\hspace{-0.1cm}
\begin{picture}(3,1.4)
\put(0.8,0.2){\vector(1,0){0.1}}
\put(2.3,0.2){\vector(1,0){0.1}}
\put(2.3,1.2){\vector(1,0){0.1}}
\put(0.8,1.2){\vector(1,0){0.1}}
\put(0.5,0.2){\line(1,0){2}}
\put(1.5,0.2){\line(1,2){0.2}}
\put(2,1.2){\line(-1,-2){0.2}}
\put(0.5,1.2){\line(1,0){2}}
\put(1,0.2){\line(0,1){1}}
\put(1.5,1.2){\line(1,-2){0.5}}
\linethickness{0.50mm}
\put(2.0,1.2){\line(1,0){0.5}}
\put(2.0,0.2){\line(1,0){0.5}}
\thinlines
\put(0.45,1.2){\makebox(0,0)[r]{$#1$}}
\put(0.45,0.2){\makebox(0,0)[r]{$#2$}}
\put(2.55,1.2){\makebox(0,0)[l]{$#3$}}
\put(2.55,0.2){\makebox(0,0)[l]{$#4$}}
\put(2.085,0.7){\makebox(0,0)[l]{$_{(l-p_1)^2}$}}
\end{picture}
}} 
\hfill}

\newcommand{\doublecrossB}[4]{
\mbox{\parbox{3.5cm}{\hspace{-0.1cm}
\begin{picture}(3,1.4)
\put(0.8,0.2){\vector(-1,0){0.1}}
\put(2.3,0.2){\vector(-1,0){0.1}}
\put(2.3,1.2){\vector(1,0){0.1}}
\put(0.8,1.2){\vector(1,0){0.1}}
\put(0.5,0.2){\line(1,0){2}}
\put(1.5,0.2){\line(1,2){0.2}}
\put(2,1.2){\line(-1,-2){0.2}}
\put(0.5,1.2){\line(1,0){2}}
\put(1,0.2){\line(0,1){1}}
\put(1.5,1.2){\line(1,-2){0.5}}
\linethickness{0.50mm}
\put(2.0,1.2){\line(1,0){0.5}}
\put(0.5,0.2){\line(1,0){0.5}}
\thinlines
\put(2.0,0.2){\line(1,0){0.5}}
\put(0.45,1.2){\makebox(0,0)[r]{$#1$}}
\put(0.45,0.2){\makebox(0,0)[r]{$#2$}}
\put(2.55,1.2){\makebox(0,0)[l]{$#3$}}
\put(2.55,0.2){\makebox(0,0)[l]{$#4$}}
\end{picture}
}} 
\hfill}

\newcommand{\doublecrossBb}[4]{
\mbox{\parbox{3.5cm}{\hspace{-0.1cm}
\begin{picture}(3,1.4)
\put(0.8,0.2){\vector(-1,0){0.1}}
\put(2.3,0.2){\vector(-1,0){0.1}}
\put(2.3,1.2){\vector(1,0){0.1}}
\put(0.8,1.2){\vector(1,0){0.1}}
\put(0.5,0.2){\line(1,0){2}}
\put(1.5,0.2){\line(1,2){0.2}}
\put(2,1.2){\line(-1,-2){0.2}}
\put(0.5,1.2){\line(1,0){2}}
\put(1,0.2){\line(0,1){1}}
\put(1.5,1.2){\line(1,-2){0.5}}
\linethickness{0.50mm}
\put(2.0,1.2){\line(1,0){0.5}}
\put(0.5,0.2){\line(1,0){0.5}}
\thinlines
\put(2.0,0.2){\line(1,0){0.5}}
\put(0.45,1.2){\makebox(0,0)[r]{$#1$}}
\put(0.45,0.2){\makebox(0,0)[r]{$#2$}}
\put(2.55,1.2){\makebox(0,0)[l]{$#3$}}
\put(2.55,0.2){\makebox(0,0)[l]{$#4$}}
\put(2.085,0.7){\makebox(0,0)[l]{$_{(k)^2}$}}
\end{picture}
}} 
\hfill}

\newcommand{\doublecrossCa}[4]{
\mbox{\parbox{3.5cm}{\hspace{-0.1cm}
\begin{picture}(3,1.4)
\put(0.7,0.2){\vector(1,0){0.1}}
\put(2.3,0.2){\vector(1,0){0.1}}
\put(2.3,1.2){\vector(1,0){0.1}}
\put(0.8,1.2){\vector(1,0){0.1}}
\put(0.5,0.2){\line(1,0){2}}
\put(1,0.2){\line(1,2){0.2}}
\put(1.5,1.2){\line(-1,-2){0.2}}
\put(0.5,1.2){\line(1,0){2}}
\put(2,0.2){\line(0,1){1}}
\put(1,1.2){\line(1,-2){0.5}}
\linethickness{0.50mm}
\put(2.0,1.2){\line(1,0){0.5}}
\put(2.0,0.2){\line(1,0){0.5}}
\thinlines
\put(0.45,1.2){\makebox(0,0)[r]{$#1$}}
\put(0.45,0.2){\makebox(0,0)[r]{$#2$}}
\put(2.55,1.2){\makebox(0,0)[l]{$#3$}}
\put(2.55,0.2){\makebox(0,0)[l]{$#4$}}
\end{picture}
}} 
\hfill}

\newcommand{\doublecrossCb}[4]{
\mbox{\parbox{3.5cm}{\hspace{-0.1cm}
\begin{picture}(3,1.4)
\put(0.7,0.2){\vector(1,0){0.1}}
\put(2.3,0.2){\vector(1,0){0.1}}
\put(2.3,1.2){\vector(1,0){0.1}}
\put(0.8,1.2){\vector(1,0){0.1}}
\put(0.5,0.2){\line(1,0){2}}
\put(1,0.2){\line(1,2){0.2}}
\put(1.5,1.2){\line(-1,-2){0.2}}
\put(0.5,1.2){\line(1,0){2}}
\put(2,0.2){\line(0,1){1}}
\put(1,1.2){\line(1,-2){0.5}}
\linethickness{0.50mm}
\put(2.0,1.2){\line(1,0){0.5}}
\put(2.0,0.2){\line(1,0){0.5}}
\thinlines
\put(0.45,1.2){\makebox(0,0)[r]{$#1$}}
\put(0.45,0.2){\makebox(0,0)[r]{$#2$}}
\put(2.55,1.2){\makebox(0,0)[l]{$#3$}}
\put(2.55,0.2){\makebox(0,0)[l]{$#4$}}
\put(2.085,0.7){\makebox(0,0)[l]{$_{(k)^2}$}}
\end{picture}
}} 
\hfill}

\newcommand{\doublecrossCc}[4]{
\mbox{\parbox{3.5cm}{\hspace{-0.1cm}
\begin{picture}(3,1.4)
\put(0.7,0.2){\vector(1,0){0.1}}
\put(2.3,0.2){\vector(1,0){0.1}}
\put(2.3,1.2){\vector(1,0){0.1}}
\put(0.8,1.2){\vector(1,0){0.1}}
\put(0.5,0.2){\line(1,0){2}}
\put(1,0.2){\line(1,2){0.2}}
\put(1.5,1.2){\line(-1,-2){0.2}}
\put(0.5,1.2){\line(1,0){2}}
\put(2,0.2){\line(0,1){1}}
\put(1,1.2){\line(1,-2){0.5}}
\linethickness{0.50mm}
\put(2.0,1.2){\line(1,0){0.5}}
\put(2.0,0.2){\line(1,0){0.5}}
\thinlines
\put(0.45,1.2){\makebox(0,0)[r]{$#1$}}
\put(0.45,0.2){\makebox(0,0)[r]{$#2$}}
\put(2.55,1.2){\makebox(0,0)[l]{$#3$}}
\put(2.55,0.2){\makebox(0,0)[l]{$#4$}}
\put(2.085,0.85){\makebox(0,0)[l]{$_{(k)^2}$}}
\put(2.085,0.55){\makebox(0,0)[l]{$_{(l-p_{123})^2}$}}
\end{picture}
}} 
\hfill}

\begin{document}
\unitlength1cm
\maketitle
\allowdisplaybreaks

\section{Introduction} 
\labbel{sec:intro}

Precision studies of the electroweak interaction at the LHC are based on a wealth of 
observables derived from vector boson pair production, 
$\gamma\gamma$, $Z\gamma$, $W\gamma$, $ZZ$, $WW$, $WZ$, 
which allow to test the $SU(2)_L\times U(1)_Y$ gauge
structure and the field content of the Standard Model.
Anomalous contributions to these interactions
can probe physics beyond the Standard Model at energy scales well beyond direct 
searches. To fully exploit these observables, precise theoretical predictions 
are of crucial importance, including especially higher order  perturbative corrections. To 
reach an accuracy in the per-cent range, thus matching the experimental 
precision at the LHC,  corrections to next-to-leading order (NLO) in the electroweak theory and 
to next-to-next-to-leading order (NNLO) in QCD are to be included. 

The full set of NLO QCD corrections~\cite{Ohnemus:1992jn,Baur:1993ir,Baur:1997kz,Dixon:1998py} 
and large parts of the NLO electroweak 
corrections~\cite{Accomando:2004de,Accomando:2005xp,Accomando:2005ra,Baglio:2013toa,Bierweiler:2013dja,Billoni:2013aba} 
have been derived for vector boson pair production. NNLO QCD 
corrections were calculated up to now for $\gamma\gamma$~\cite{Catani:2011qz} and 
$Z\gamma$~\cite{Grazzini:2013bna} production. Key ingredient to the NNLO calculations are the 
two-loop matrix elements for $q\bar q \to V_1V_2$, which are known for 
$\gamma \gamma$~\cite{Bern:2001df,Anastasiou:2002zn} and 
$V\gamma$~\cite{Gehrmann:2011ab,Gehrmann:2013vga} production, and for
$q\bar q\to WW$ in the high energy approximation~\cite{Chachamis:2008yb}. 
The full calculation of two-loop matrix 
elements for the production of two massive vector bosons is still an outstanding task, and 
requires the derivation of a new class of two-loop Feynman integrals: two-loop four-point 
functions with internal massless propagators and two massive external legs. First results 
on these were obtained already, with the derivation of the full set of 
planar two-loop integrals for vector boson pairs with equal mass~\cite{Gehrmann:2013cxs} and 
two different masses~\cite{Henn:2014lfa}. 
In the present paper, we extend our earlier calculation~\cite{Gehrmann:2013cxs} 
to compute the full set of 
two-loop four point functions relevant to vector boson pair production with two equal 
masses. A subset of these, the three-point functions with three massive external
legs and all massless internal propagators, has already been known for some
time in the literature~\cite{Birthwright:2004kk,Chavez:2012kn}.

The problem kinematics and notation are described in section~\ref{sec:notation}. 
Working in dimensional regularization with $d=4-2\epsilon$ 
space-time dimensions, we identify the relevant master
integrals (MI) and derive differential equations for them employing
integration-by-parts~(IBP)~\cite{Tkachov:1981wb,Chetyrkin:1981qh}
and Lorentz-invariance~(LI)~\cite{Gehrmann:1999as} reductions
through the Laporta algorithm~\cite{Laporta:2001dd} implemented in the
Reduze code~\cite{Studerus:2009ye,vonManteuffel:2012np}.
The master integrals are then determined by solving these differential 
equations~\cite{Kotikov:1990kg,Remiddi:1997ny,Caffo:1998du,Gehrmann:1999as}
and matching generic solutions to appropriate boundary values obtained
in special kinematical limits.
Improving upon our earlier results~\cite{Gehrmann:2013cxs}, we are now 
transforming the differential equations to a canonical form~\cite{Henn:2013pwa}
which renders their integration trivial after an expansion in $\epsilon$.
The algorithm applied for this transformation is described in detail in 
section~\ref{sec:basis}, similar procedures have been put forward most recently 
in~\cite{Argeri:2014qva,Caron-Huot:2014lda}. With this, the remaining non-trivial step in the
calculation of the master integrals is the determination of the boundary terms,
which we describe in section~\ref{sec:boundaries}.
We use a similar setup for our parametrization and treatment of functions
as in the first calculation~\cite{vonManteuffel:2013uoa} of non-planar double
boxes with this type of external kinematics.
In particular, our solutions are described in terms of multiple polylgarithms.
We fix the boundary terms of all complicated integrals by imposing
a simple set of regularity conditions.
The implementation of these conditions and further processing of the
multiple polylogarithms relies on computer-algebra implementations of
the coproduct augmented symbol formalism~\cite{Brown:2008um,Goncharov:2010jf,Duhr:2011zq,Duhr:2012fh,Anastasiou:2013srw}
and other techniques, which we described in~\cite{Gehrmann:2013cxs,Bonciani:2013ywa}.
Section~\ref{sec:checks} contains a discussion of our solutions
and the checks we performed on them.
In section~\ref{sec:realfuncs} we describe the final form of our
analytical results in terms of a particular set of real valued $\Li_{2,2}$,
$\Li_n$ $(n=2,3,4)$ and $\ln$ functions, optimized for fast and stable
numerical evaluations.
We complement our exact results by expanding them at the production threshold
and in the high-energy region.
We conclude in section~\ref{sec:conc} and specify the exact definition
of our canonical basis in appendix~\ref{app:basis}.
For all algebraic manipulations we made extensive use of FORM~\cite{Vermaseren:2000nd}
and Mathematica~\cite{mathematica8}.

\section{Notation and reduction to master integrals}
\labbel{sec:notation}
We consider the production of two vector bosons $V$
of mass $m$ in the scattering kinematics:
\begin{equation}
 q(p_1) + \bar q(p_2) \to V(q_1) + V(q_2)\,,
\end{equation}
where $p_1^2 = p_2^2 =0$ and $q_1^2=q_2^2=m^2\,.$
The Mandelstam invariants are 
\begin{equation}
 s=(p_1+p_2)^2\,, \quad t = (p_1-q_1)^2\,, \quad u = (p_2-q_1)^2\,,
 \quad\mbox{with}\quad s+t+u = 2m^2\,,
\end{equation}
so that in the physical region relevant for vector-boson pair production
we have:
\begin{equation}
 s>4m^2\,, \quad t<0\,,\quad u<0\,,\quad \mbox{with}\quad m^2>0\,.
\end{equation}
We choose to work with dimensionless variables $x$, $y$ and $z$ defined by
\begin{equation}
\label{eq:xyz}
s = m^2 \frac{(1+x)^2}{x}\,,\quad t = - m^2 y\,, \quad u = - m^2 z\,,
\quad
\mbox{with}\quad
\frac{(1+x)^2}{x} - y - z = 2
\end{equation}
The Landau variable $x$ absorbs a square root $\sqrt{s(s-4m^2)}$ in the
differential equations which is associated with the two massive particle
threshold.

Following the work started in~\cite{Gehrmann:2013cxs} 
we organize all Feynman integrals required for the computation of $q\bar{q}\to VV$
into three different integral families named {\it Topo~A}, {\it Topo~B} and
{\it Topo~C},
where the first two topologies are needed to represent respectively 
the double-boxes with adjacent and non-adjacent massive legs,
while the third contains all non-planar integrals.
We choose the propagators of the three topologies as listed in Table~\ref{tab:auxtopo}.

\begin{table}[h!]
\begin{center}
\begin{tabular}[H]{lll}
{\it Topo~A}\hspace{5mm} & {\it Topo~B}\hspace{5mm} & {\it Topo~C}\\[2mm]
$k^2$               & $k^2$               & $k^2$               \\
$l^2$               & $l^2$               & $l^2$               \\
$(k-l)^2$           & $(k-l)^2$           & $(k-l)^2$           \\
$(k-p_1)^2$         & $(k-p_1)^2$         & $(k-p_1)^2$         \\
$(l-p_1)^2$         & $(l-p_1)^2$         & $(l-p_1)^2$         \\
$(k-p_1-p_2)^2$     & $(k-p_1+q_1)^2$     & $(k-p_1-p_2)^2$     \\
$(l-p_1-p_2)^2$     & $(l-p_1+q_1)^2$     & $(k-l  -q_1)^2$     \\
$(k-p_1-p_2+q_1)^2$ & $(k-p_1-p_2+q_1)^2$ & $(l-p_1-p_2+q_1)^2$ \\
$(l-p_1-p_2+q_1)^2$ & $(l-p_1-p_2+q_1)^2$ & $(k-l-p_1-p_2)^2$  
\end{tabular}
\end{center}
\caption{Propagators in the three different integral families used 
to represent all two-loop 4-point integrals 
with two massless and two massive legs with the same mass.}
\labbel{tab:auxtopo}
\end{table}

As it is well known, using IBPs, LIs and symmetry relations all 
Feynman integrals described by these three integral 
families can be reduced to a small subset, the master integrals.
We performed this reduction for all integrals relevant for our process
using the automated codes 
Reduze\;1 and Reduze\;2~\cite{Studerus:2009ye,vonManteuffel:2012np,Bauer:2000cp,fermat}.
After the reduction we find that all integrals can be expressed in
terms of $75$ MIs, some of which are actually not 
genuinely independent, but can instead be
related to each other through a permutation of the external legs $p_1 \leftrightarrow p_2$. 

In~\cite{Gehrmann:2013cxs}, we described 
the computation of the MIs embedded in {\it Topo~A} and {\it Topo~B}. 
In the present work we conclude the computation of all 
non-planar MIs in {\it Topo~C}.

\section{Building up a canonical basis}
\labbel{sec:basis}
It has been recently suggested~\cite{Henn:2013pwa} that a suitable choice of basis for the 
MIs renders their computation in the differential equation method more transparent. 
In particular it has been conjectured that, 
if all MIs for a given topology can be integrated in terms of Chen iterated integrals 
only~\cite{Chen:1977oja,Henn:2013woa} (the
commonly used generalized 
harmonic polylogarithms, GHPLs, are a special case of 
these~\cite{Remiddi:1999ew,Gehrmann:2000zt,Zagier,Goncharov,2001math.3059G,Goncharov:2005sla}), then there
must exist a basis choice, with $\vec{m} = \{\,m_j\,\}$, such that the differential equations 
with respect to the external invariants can be cast in the \textsl{canonical form}:

\begin{equation}
d\,\vec{m}(x_j; \epsilon) = \epsilon\, d\,A(x_j)\, \vec{m}(x_j; \epsilon)\,, \labbel{eq:cform}
\end{equation}
where $x_j$ are the external invariants, 
the differential $d$ acts on all external invariants, and the dependence on the
dimensional regularization parameter $\epsilon$ is completely factorized from the
kinematics.

Moreover, if the matrix $A(x_j)$ can be written as:
\begin{equation}
A(x_j) = \sum_{k=1}^n\, A_k\, \ln{r_k}\,, \labbel{eq:Adlog}
\end{equation}
where $A_k$ are \textsl{constant} matrices and $r_k$ are simple \textsl{rational functions}
of the external invariants $x_j$ then, by their very definitions, upon integration of~\eqref{eq:cform},
the result will only contain GHPLs of alphabet $\{\,r_k\,\}_{k=1}^n.$

While casting the differential equations in the canonical form~\eqref{eq:cform} 
is not strictly necessary for their integration, it is still very desirable 
for different reasons. 
In particular the MIs computed in the canonical basis $\vm$, once expanded as
Laurent series in $\epsilon$,
end up having a particularly compact representation in terms of \textsl{pure functions
of uniform transcendentality}, i.e. order by order in $\epsilon$ each MI is given only
by a combination of transcendental functions of uniform weight~\cite{Henn:2013pwa}.
Having a result in this form, in particular in the case of multi-scale and/or multi-loop
problems, helps to handle the largeness of the intermediate expressions 
and also the complexity of the final result. 

The issue of the existence of such a basis for any multi-loop problem remains,
in particular for those cases which cannot be expressed in terms of GHPLs or
general Chen Iterated integrals~\cite{Laporta:2004rb,Adams:2013nia,Remiddi:2013joa,Bloch:2013tra}.
Moreover, even in those cases where it is known that the final result will contain only
GHPLs, no algorithm for finding such basis is known, while only some general
criteria have been 
pointed out  recently~\cite{Henn:2013tua,Henn:2013nsa,Argeri:2014qva,Caron-Huot:2014lda}.

In what follows we will describe in detail the procedure that we exploited in order
to build up a canonical basis in the explicit case under study.
While we claim no generality in this approach and no proof can be given that such approach
would work in more involved cases, we found it particularly elementary and 
algorithmically straightforward to implement, so that its extension to more difficult
cases should not present particular conceptual difficulties.

\subsection{Building up the basis bottom-up in $t$}

Before describing in detail the method we used to find our canonical basis,
let us recall some notation and definitions which will be useful in the following.
We start off by considering a \textsl{topology} (or \textsl{sector})
given by a set of $t$ different propagators (matching the loop integrals
of some Feynman diagram).
Its \textsl{sub-topologies} (or \textsl{sub-sectors}) are defined as the set of
all possible arrangements of propagators obtained from the original topology
by removing one or more propagators in all possible ways.
In the case of two-loop corrections to vector boson pair production in massless QCD, 
where all tadpoles are identically zero,
the first non-zero sectors will be those with $t=3$  
(corresponding to the \textsl{sunrise topologies}), while the highest sectors
will contain at most $t=7$ different propagators.

As it is well known, by generating and solving all IBPs, LIs and symmetry relations for a
given integral family, some sectors will be reduced to one or more MIs while some other,
the so-called \textsl{reducible sectors}, will be completely reduced to their sub-topologies.
In what follows we can completely neglect these reducible sectors.

Let us consider now a sector with a given value of $t$ and which is reduced
to $n \geq 1$ MIs. As it is well known, the differential equations for the latter
will in general contain all their sub-topologies as inhomogeneous terms.
Therefore, it is natural to try and follow a bottom-up approach in $t$,
such that, when studying the differential equations for the MIs of 
a given sector with a given value of $t$, we can assume that all its
sub-topologies fulfil differential equations already in canonical form.
For the MIs of the sector under consideration we use the notation: 
$f_a(x_k;\epsilon)$  with $a=1,...,n$. 
The differential equations for the $n$ MIs read in total generality:
\begin{equation}
\frac{\partial}{\partial x_j} \, f_a(x_k;\epsilon) = C_{ab}^{(j)}(x_k;\epsilon)\,f_b(x_k;\epsilon)
+ D_{al}^{(j)}(x_k;\epsilon)\, m_l(x_k;\epsilon)\,, \labbel{eq:deq}
\end{equation}
where the $C_{ab}^{(j)}(x_k;\epsilon)$ and the $D_{al}^{(j)}(x_k;\epsilon)$ are 
at most rational functions of $x_k$ and $\epsilon$, while the $m_l(x_k;\epsilon)$
are the sub-topologies, whose differential equations are, by construction, already
in the canonical form:

\begin{equation}
 \frac{\partial}{\partial x_j} \, m_l \, (x_k;\epsilon) 
 = \epsilon \, A_{lr}^{(j)}(x_k) \, m_r(x_k;\epsilon)\,.\labbel{eq:deqsub}
\end{equation}

Let us consider now the $n \times n$ matrix of the coefficients of the homogeneous
equation $\C(x_k;\epsilon) = \{\,C_{ab}^{(j)}(x_k;\epsilon)\,\}_{ab}$. 
Our method relies on the assumption that we can find 
a starting basis of MIs $f_a(x_k;\epsilon)$ such that:

\begin{enumerate}

\item The matrix $\C(x_k,\epsilon)$ has only \textsl{linear dependence}\footnote{Note 
that this same requirement was assumed in~\cite{Argeri:2014qva}, but for the \textsl{entire}
system of differential equations for \textsl{all MIs}, including all sub-topologies, while
here we require it, for every value of $t$, only for the homogeneous part of the system.} 
 on $\epsilon$, i.e
      \begin{equation}
       \C(x_k,\epsilon) = \C^{(0)}(x_k) + \epsilon\; \C^{(1)}(x_k)\,.\labbel{eq:lin}
      \end{equation}

      \item The matrix $\C^{(0)}(x_k)$ is \textsl{triangular}.
\end{enumerate}

Obviously in the case where $n=1$ the matrix $\C(x_k;\epsilon)$ reduces to a \textsl{scalar}
and the condition $2.$ is always trivially satisfied. 
On the other hand there is no real restriction on the dependence on $\epsilon$
of the functions $D_{al}^{(j)}(x_k;\epsilon)$. 
As an exemplification
we can assume a typical situation where they contain terms of the following form:

\begin{equation}
 D_{al}^{(j)}(x_k;\epsilon) = \alpha_{al}^{(j)}(x_k) 
 + \beta_{al}^{(j)}(x_k) \; \epsilon 
 + \frac{\gamma_{al}^{(j)}(x_k)}{1-2\,\epsilon}
\,, \labbel{eq:Deps}
\end{equation}
where the functions $\alpha_{al}^{(j)}$, $\beta_{al}^{(j)}$, $\gamma_{al}^{(j)}$
depend only on the external invariants $x_k$.
Note that if the factor $1/(1-2\epsilon)$ were substituted by any other linear
factor $1/(u+v\,\epsilon)$, with $u,v \in \Z$, the argument would proceed in the
exact same way.
Moreover, as it will become clear in what follows, a more complicated dependence on
$\epsilon$ in the inhomogeneous terms (for example polynomial in $\epsilon$) 
can be, at least in principle, treated with a 
suitable extension of the method described below.\newline

For every sector at a given $t$ we proceeded as follows:

\begin{enumerate}
 \item Starting from~\eqref{eq:deq}, and using the assumption~\eqref{eq:lin},
 we first attempt to solve the homogeneous system for $\epsilon = 0$
 \begin{equation}
  \frac{\partial}{\partial x_j} \, \vf(x_k) = C^{(0)}(x_k)\,\vf(x_k)\,, \quad \forall\; j\;,
 \end{equation}
 in terms of \textsl{rational functions only}. 
 While there is obviously \textsl{a priori} no guarantee that this can
 be done in general (without introducing, for example, 
 logarithms of the external invariants $x_k$), we found that, in all cases we 
 worked with, this
 was always the case. If this is possible, then it is equivalent to finding a
 rotation $f_a(x_k;\epsilon) \to g_a(x_k;\epsilon)$ such that the system~\eqref{eq:deq}
 becomes:
 
 \begin{equation}
\frac{\partial\,g_a(x_k;\epsilon)}{\partial x_j}  = 
\epsilon\;\widetilde{C}_{ab}^{(j)}(x_k)\,g_b(x_k;\epsilon)
+ \widetilde{D}_{al}^{(j)}(x_k;\epsilon)\, m_l(x_k;\epsilon)\,, \labbel{eq:deqb}
\end{equation}

where the functions $\widetilde{C}_{ab}^{(j)}(x_k)$ are simple rational functions
of the external invariants only, while the $\widetilde{D}_{al}^{(j)}(x_k;\epsilon)$
will have in general the same decomposition as in~\eqref{eq:Deps}:
\begin{equation}
 \widetilde{D}_{al}^{(j)}(x_k;\epsilon) = \tilde{\alpha}_{al}^{(j)}(x_k) 
 + \tilde{\beta}_{al}^{(j)}(x_k) \; \epsilon 
 + \frac{\tilde{\gamma}_{al}^{(j)}(x_k)}{1-2\,\epsilon}
\,. \labbel{eq:Depsb}
\end{equation}
 
 \item Once the differential equations are in form~\eqref{eq:deqb}, only the sub-topologies
 need to be fixed in order to achieve a complete canonical form. Assuming an $\epsilon$-dependence
 as in eq.\eqref{eq:Depsb}, we start removing \textsl{first} all subtopologies
 proportional to the coefficients $\gamma_{al}^{(j)}$.
 This can be attempted performing a shift in
 the MIs basis as follows:
 
 \begin{align}
  g_a(x_k;\epsilon) \;\to\; h_a(x_k;\epsilon) = g_a(x_k;\epsilon) 
                       + \frac{\Gamma_{al}(x_k)}{1-2\,\epsilon} \,m_l(x_k,\epsilon)\,,
                       \labbel{eq:shifta}
 \end{align}
 where the $\Gamma_{al}(x_k)$ are rational functions of the external invariants
 and whose explicit form will be determined in the following.
 Note that, since the differential equations for the sub-topologies are already in 
 canonical form~\eqref{eq:deqsub}, this ensures that upon performing this shift
 and partial-fractioning in $\epsilon$
 we will only produce terms proportional $\epsilon^0$, $\epsilon$,
 or $1/(1-2\epsilon)$.
 Upon performing the shifts in~\eqref{eq:shifta}, in fact, we are left with:
 
 \begin{align}
  \frac{\partial\,h_a(x_k;\epsilon)}{\partial\,x_j} 
  &= \epsilon\,\widetilde{C}_{ab}^{(j)}(x_k)\,h_{b}(x_k;\epsilon)
  + \epsilon \, \tilde{\beta}_{al}^{(j)}(x_k)\,m_l(x_k;\epsilon)\nonumber \\
  &+\left[\tilde{\alpha}_{al}^{(j)}(x_k) 
           + \frac{\Gamma_{bl}(x_k)}{2}\, 
               \widetilde{C}_{ab}^{(j)}(x_k) \right] m_l(x_k;\epsilon)\nonumber \\
  &-\frac{1}{2} \Gamma_{al}(x_k)\,A_{lr}^{(j)}(x_k)\,m_r(x_k;\epsilon)\nonumber \\
  &+\frac{1}{1-2\epsilon}\Bigg\{
                         \left[ \frac{\partial \Gamma_{al}(x_k)}{\partial x_j} 
                              + \tilde{\gamma}_{al}^{(j)}(x_k) 
             - \frac{1}{2} \widetilde{C}_{ab}^{(j)}(x_k)\,\Gamma_{bl}(x_k) \right] m_l(x_k;\epsilon)\nonumber\\
  &\hspace{2cm} + \frac{1}{2}\, \Gamma_{al}(x_k)\,A_{lr}^{(j)}(x_k)\,m_r(x_k;\epsilon)\,\Bigg\}\,.
  \labbel{eq:deqshifted}
  \end{align}

The explicit form of the functions $\Gamma_{al}(x_k)$
can be, at least in principle, determined by imposing that all terms 
proportional to $1/(1-2\epsilon)$ are cancelled, in other words that:

\begin{align}
 &\left[ \frac{\partial \Gamma_{al}(x_k)}{\partial x_j} 
                              + \tilde{\gamma}_{al}^{(j)}(x_k) 
             - \frac{1}{2} \widetilde{C}_{ab}^{(j)}(x_k)\,\Gamma_{bl}(x_k) \right] m_l(x_k;\epsilon)
             \nonumber \\
  &\hspace{2cm} + \frac{1}{2}\, \Gamma_{al}(x_k)\,A_{lr}^{(j)}(x_k)\,m_r(x_k;\epsilon) = 0\,.
 \labbel{eq:eqan}
\end{align}

Eq.~\eqref{eq:eqan} is a linear system of first-order coupled differential equations 
for the unknown $\Gamma_{al}(x_k)$ whose solution can be,
at least in principle, as difficult as the solution of the original
system~\eqref{eq:deq}. Nevertheless, in all cases that we encountered, 
the system could be easily solved with an \textsl{Ansatz}.
In particular, assuming that a basis which realizes the canonical 
form~\eqref{eq:cform} exists and assuming that such basis 
can be reached through a rotation which
only involves \textsl{rational functions}\footnote{Note that
this requirement is perfectly sensible as long as we assume that such basis
can be reached from any other basis through IBPs, LIs and symmetry relations only and
potential roots coming from the solution of the homogeneous equations 
can be rationalized, similar like in our case where the
Landau variable $x$ absorbs the root $\sqrt{s(s-4m^2)}$.}, 
we can write the most general
\textsl{Ansatz} for the functions $\Gamma_{al}(x_k)$ as linear combination of
all possible linearly independent rational functions\footnote{Where
here \textsl{``linearly independent''} has to be intended in the sense of
a complete partial fractioning in all external invariants $x_k$.} which 
appear in the original differential equations~\eqref{eq:deqb}.
Collecting for the independent rational functions, and requiring their coefficients
to be zero, we are left with a large system of linear
equations with numerical coefficients whose solution is now,
at least in principle, completely straightforward.

Note that, in a typical case, there will be more equations than unknowns
and the system will be over-constrained with many equations
being linearly dependent from each other. For the same reason,
it is in no way guaranteed that a solution to such a system exists. 
Nevertheless, once more, for all cases that we worked with, a solution
could always be found.

\item Once this step has been performed, we are left with a new system of
equations for the new MIs $h_a(x_k;\epsilon)$ which reads:

\begin{equation}
 \frac{\partial}{\partial x_j} \, h_a(x_k;\epsilon) = 
 \epsilon \,\left[ \, \widetilde{C}_{ab}^{(j)}(x_k)\,h_b(x_k;\epsilon)
 + E_{al}^{(j)}(x_k)\,m_l(x_k;\epsilon)\,\right]
 \;+\; F_{al}^{(j)}(x_k)\, m_l(x_k;\epsilon)\,, \labbel{eq:deqc}
\end{equation}

where the $E_{al}^{(j)}(x_k)$ and the $F_{al}^{(j)}(x_k)$ are again simple
rational functions of the external invariants $x_k$. 
We can now proceed removing the
remaining terms which are not proportional to $\epsilon$. 
This can be achieved in the same way as before by performing the shift:
\begin{equation}
 h_a(x_k;\epsilon) \to m_a(x_k;\epsilon) = h_a(x_k;\epsilon) 
 + G_{al}(x_k)\,m_l(x_k;\epsilon)\,,
 \labbel{eq:shiftb}
\end{equation}

where again the $G_{al}(x_k)$ are rational functions of the external invariants.

Note that, since the MIs depend in general on many external invariants $x_j$,
for every $a,l$ fixed, there has to exist a single function $G_{al}(x_k)$, such that
the terms not proportional to $\epsilon$ in~\eqref{eq:deqc} cancel under
the shift~\eqref{eq:shiftb} for all differential equations in all external invariants $x_j$.
This condition can be rephrased as:

\begin{equation}
 \frac{\partial}{\partial\,x_j} G_{al}(x_k) + F_{al}^{(j)}(x_k) = 0\,, \qquad \forall\;j. 
\end{equation}

With the same assumptions as before we can solve these equations with an \textsl{Ansatz}
imposing that the solution must be a linear combination of rational functions
in the external invariants only. Again, in all cases where we applied this method,
a solution could always be found.

\item After the final shift~\eqref{eq:shiftb} is performed the canonical form is reached:

\begin{equation}
 \frac{\partial \,m_a(x_k;\epsilon)}{\partial x_j}  = 
 \epsilon \, \, H_{ab}^{(j)}(x_k)\,m_b(x_k;\epsilon)\,, \labbel{eq:deqfin}
\end{equation}
where the $H_{ab}^{(j)}(x_k)$ are only rational functions and
the indices $\{a,b\}$ run on the MIs of the given sectors plus
on all their sub-topologies.

\end{enumerate}

We would like to emphasize that 
 there is no guarantee that, given 
 any sets of MIs to start with, all steps described above can be always 
 successfully carried out. These require in each instance to find a shift which only involves
 rational functions and which eliminates at every step the un-wanted terms
 in the differential equations.
It should also be noted that the two steps $2.$ and $3.$ must be performed in this order.
 It is clear in fact from~\eqref{eq:deqshifted} that the first step will
 produce in general terms in the equations which are proportional to
 $\epsilon^0$, and which will be removed only during the following step.

\subsection{Extension to polynomial dependence on $\epsilon$}
In the very same way as discussed above we can treat the more general
case where the differential equations~\eqref{eq:deqb} 
admit a polynomial dependence on $\epsilon$ in the sub-topologies
and/or higher powers of factors $1/(u+v\,\epsilon)$.

Let us consider for simplicity a sector with only one master integral $f(x;\epsilon)$,
which depends on one single external invariant $x$ and which
has only one sub-topology $m(x;\epsilon)$. 
Let us assume again that the differential equation for the sub-topology
is in canonical form

\begin{equation}
 \frac{d}{dx} m(x;\epsilon) = \epsilon\,A(x)\,m(x;\epsilon)\,.\labbel{eq:polcan}
\end{equation}

Moreover, let us suppose that the differential equation for the MI $f(x;\epsilon)$
is \textsl{almost} in canonical form except for a term proportional to $\epsilon^n$,

\begin{equation}
 \frac{d}{dx} f(x;\epsilon) = \epsilon\,\left( C_1(x) f(x;\epsilon) + C_2(x)\,m(x;\epsilon) \right)
                            + \epsilon^n\,C_3(x)\,m(x;\epsilon)\,,\labbel{eq:poldeq}
\end{equation}
where the $C_j(x)$ are all rational functions of the external invariant $x$.

Using again~\eqref{eq:polcan}, we can perform the shift:

\begin{equation}
 f(x;\epsilon) \to g(x;\epsilon) = f(x;\epsilon) + \epsilon^{n-1}\,G(x)\,m(x;\epsilon)\,.
 \labbel{eq:shiftpol}
\end{equation}
Inserting this expression in~\eqref{eq:poldeq} and using the fact that the differential
equations for the sub-topology $m(x;\epsilon)$ are in canonical form, 
it is clear that we will produce terms proportional
to $\epsilon^n$ and $\epsilon^{n-1}$ only. 
We can then fix the function $G(x)$ 
imposing that the shift~\eqref{eq:shiftpol} removes all terms proportional to $\epsilon^n$,
being in this way left only with terms proportional to $\epsilon^{n-1}$.
Proceeding in this way, starting from the highest power of $\epsilon$, we can tentatively
remove all undesirable terms from the differential equations and bring them 
to the canonical form~\eqref{eq:cform}.

The very same idea applies to higher powers of factors $1/(u+v \,\epsilon)$ which
multiply any subtopology whose equations are already in canonical form.
Starting from the highest powers we can tentatively remove all terms one after the other
until we are reduced to the case treated in the section above.

\subsection{The basis}
\labbel{sec:thebasis}
Applying the algorithm described above we could find a canonical basis for all MIs
contributing to both planar and non-planar corrections to $q\bar{q} \to VV$.
In total there are $75$ MIs, some of which are not truly independent but instead are
related by a permutation of the external legs $p_1 \leftrightarrow p_2$
(or equivalently $q_1 \leftrightarrow q_2$).

Following the ideas described above, we started building up our basis 
with the following initial choice of MIs:

{\small
\begin{align*}
f_{1}^\topoid{A38}  &\!= \bubbleNLO{p_{12}}\!  &
f_{2}^\topoid{A134} &\!= \bubbleNLO{q_2}\!  &
f_{3}^\topoid{A148} &\!= \bubbleNLO{p_{23}}\! & 
f_{4}^\topoid{\thickbar{A}148} &\!= \bubbleNLO{p_{13}}  
\end{align*}

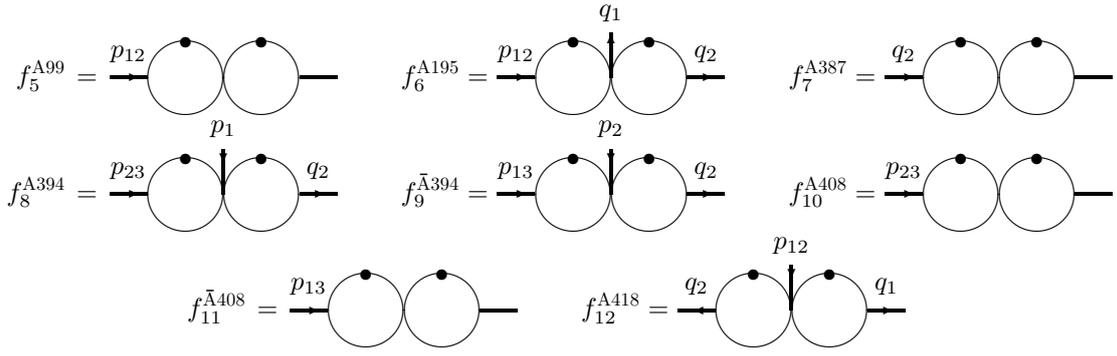
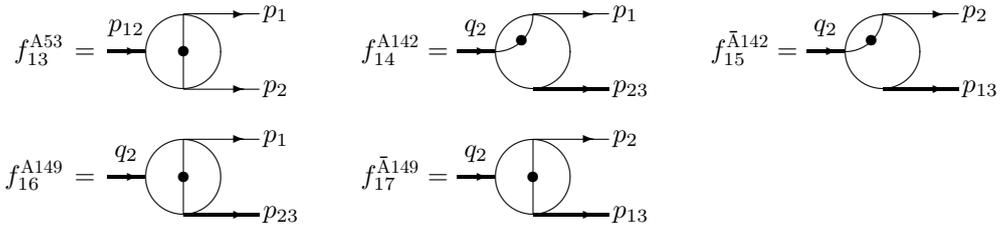
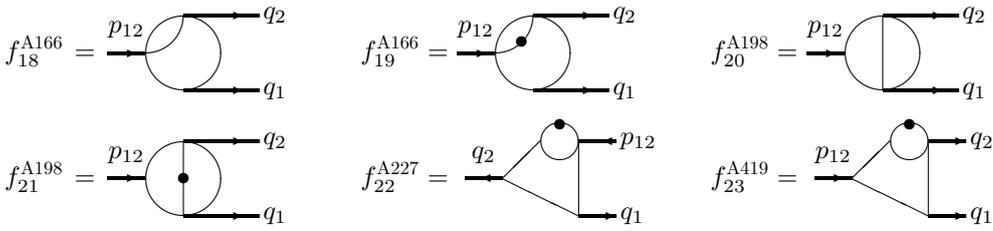
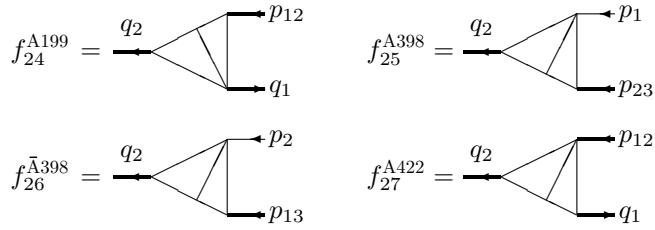
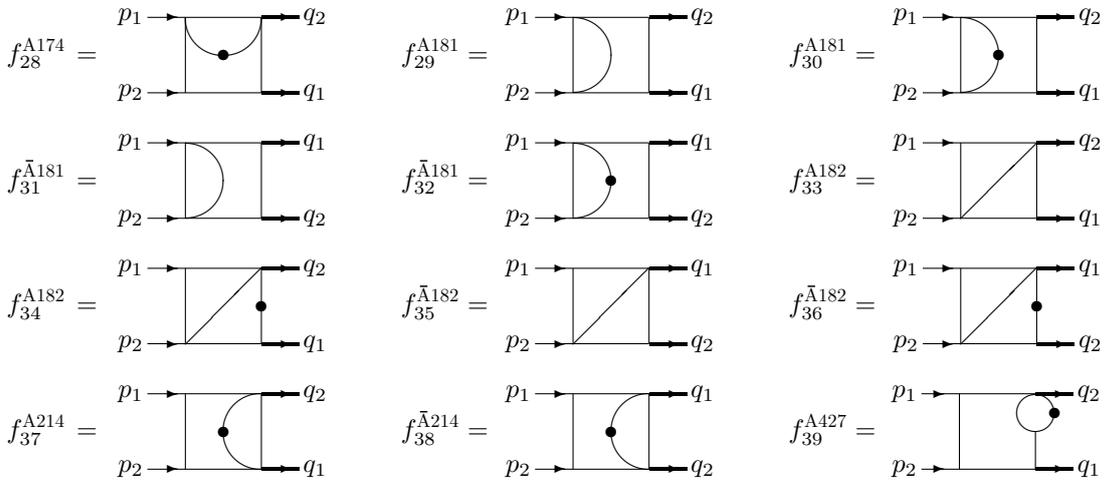
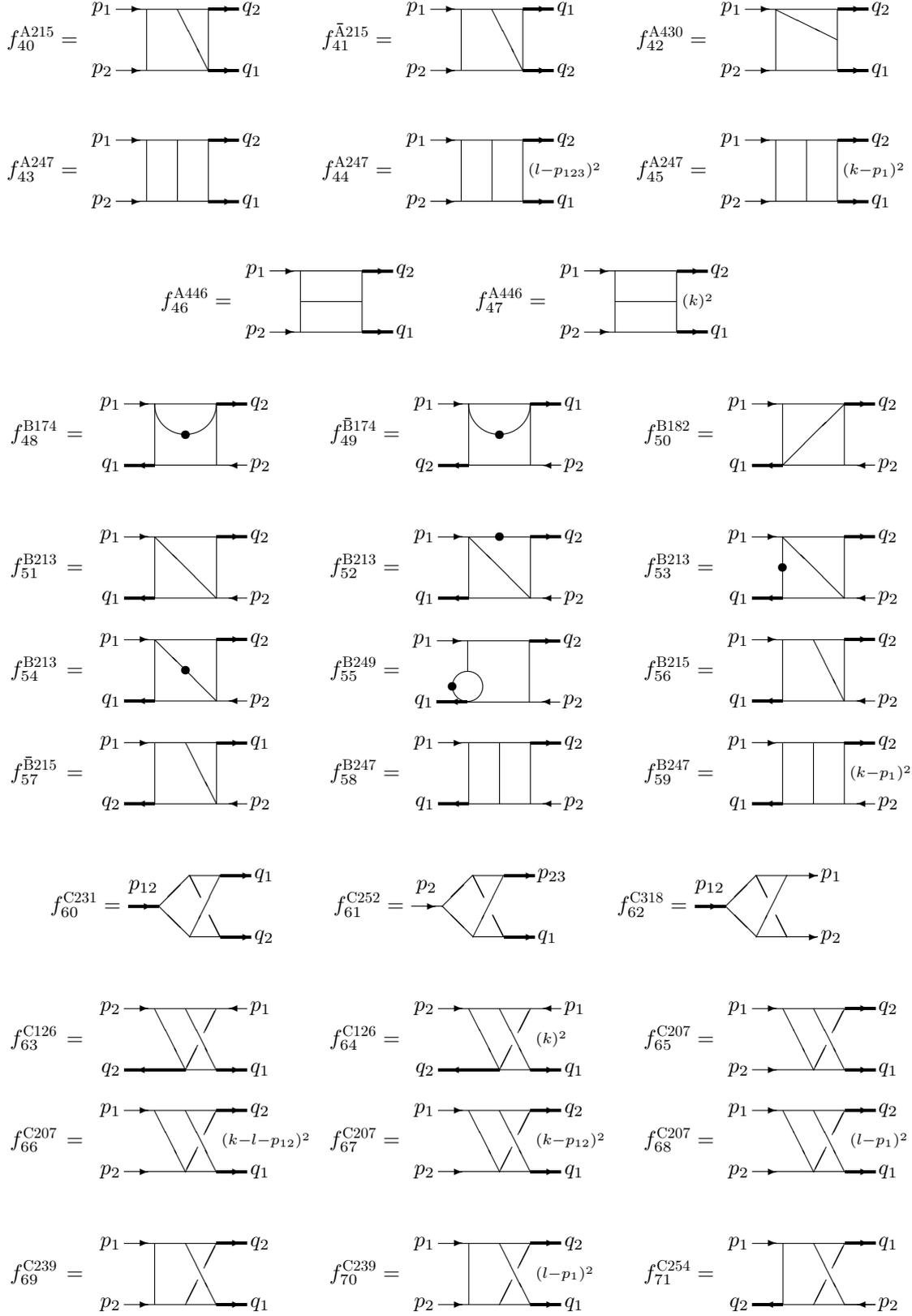
\begin{eqnarray*}
 f_{5}^\topoid{A99} &=& \doublebubble{p_{12}} \quad
 f_{6}^\topoid{A195}  = \doublebubblex{p_{12}}{q_{2}}{q_{1}} \quad
 f_{7}^\topoid{A387}  = \doublebubble{q_2} \\&&\\
 f_{8}^\topoid{A394} &=& \doublebubblexb{p_{23}}{q_{2}}{p_{1}} \quad
 f_{9}^\topoid{\thickbar{A}394}  = \doublebubblexb{p_{13}}{q_{2}}{p_{2}} \quad
 f_{10}^\topoid{A408} = \doublebubble{p_{23}}
\end{eqnarray*}
\begin{eqnarray*}
 f_{11}^\topoid{\thickbar{A}408} &=& \doublebubble{p_{13}} \quad
 f_{12}^\topoid{A418}  = \doublebubblexc{q_{2}}{q_{1}}{p_{12}} \quad
\end{eqnarray*}

\begin{eqnarray*}
f_{13}^\topoid{A53} &=& \triangleOne{p_{12}}{p_1}{p_2}  \quad
f_{14}^\topoid{A142} = \triangleTwox{q_2}{p_1}{p_{23}}  \quad
f_{15}^\topoid{\thickbar{A}142} = \triangleTwox{q_2}{p_2}{p_{13}} \\
f_{16}^\topoid{A149} &=& \triangleTwo{q_2}{p_1}{p_{23}}  \quad
f_{17}^\topoid{\thickbar{A}149} = \triangleTwo{q_2}{p_2}{p_{13}}  \quad
\end{eqnarray*}

\begin{eqnarray*}
f_{18}^\topoid{A166} &=&\triangleThreex{p_{12}}{q_2}{q_1}  \quad 
f_{19}^\topoid{A166} =\triangleThreexdot{p_{12}}{q_2}{q_1} \quad
f_{20}^\topoid{A198} = \triangleThree{p_{12}}{q_2}{q_1}   \\
f_{21}^\topoid{A198} &=& \triangleThreedot {p_{12}}{q_2}{q_1}\quad
f_{22}^\topoid{A227} = \trianglebubbleb{q_{2}}{p_{12}}{q_1} \quad
f_{23}^\topoid{A419} = \trianglebubble{p_{12}}{q_{2}}{q_1}
\end{eqnarray*}

\begin{eqnarray*}
f_{24}^\topoid{A199} &=& \trianglebThree{q_2}{p_{12}}{q_1} \quad 
f_{25}^\topoid{A398} = \trianglebTwo{q_2}{p_1}{p_{23}} \\
f_{26}^\topoid{\thickbar{A}398} &=& \trianglebTwo{q_2}{p_2}{p_{13}} \quad
f_{27}^\topoid{A422} = \trianglebThreex{q_2}{p_{12}}{q_1} 
\end{eqnarray*}

\begin{eqnarray*}
f_{28}^\topoid{A174} &=& \boxbubblea{p_1}{p_2}{q_2}{q_1}  \quad
f_{29}^\topoid{A181} = \boxbubblec{p_1}{p_2}{q_2}{q_1}  \quad
f_{30}^\topoid{A181} = \boxbubblecdot{p_1}{p_2}{q_2}{q_1} \\
f_{31}^\topoid{\thickbar{A}181} &=& \boxbubblec{p_1}{p_2}{q_1}{q_2}   \quad 
f_{32}^\topoid{\thickbar{A}181} = \boxbubblecdot{p_1}{p_2}{q_1}{q_2} \quad
f_{33}^\topoid{A182} = \boxxbNLO{p_1}{p_2}{q_2}{q_1}  \\ 
f_{34}^\topoid{A182} &=& \boxxbdotNLO{p_1}{p_2}{q_2}{q_1} \quad
f_{35}^\topoid{\thickbar{A}182} = \boxxbNLO{p_1}{p_2}{q_1}{q_2}  \quad 
f_{36}^\topoid{\thickbar{A}182} = \boxxbdotNLO{p_1}{p_2}{q_1}{q_2} \\
f_{37}^\topoid{A214} &=& \boxbubbleb{p_1}{p_2}{q_2}{q_1} \quad  
f_{38}^\topoid{\thickbar{A}214} = \boxbubbleb{p_1}{p_2}{q_1}{q_2} \quad
f_{39}^\topoid{A427} = \boxbubble{p_1}{p_2}{q_2}{q_1}
\end{eqnarray*}

\begin{eqnarray*}
f_{40}^\topoid{A215} = \boxxbpaNLO{p_1}{p_2}{q_2}{q_1} \quad 
f_{41}^\topoid{\thickbar{A}215} = \boxxbpaNLO{p_1}{p_2}{q_1}{q_2} \quad
f_{42}^\topoid{A430} = \boxxbpbNLO{p_1}{p_2}{q_2}{q_1}
\end{eqnarray*}        
\begin{eqnarray*}
f_{43}^\topoid{A247} = \doubleboxa{p_1}{p_2}{q_2}{q_1} \quad   
f_{44}^\topoid{A247} = \doubleboxatwo{p_1}{p_2}{q_2}{q_1} \quad  
f_{45}^\topoid{A247} = \doubleboxathree{p_1}{p_2}{q_2}{q_1} 
\end{eqnarray*}
\begin{eqnarray*}
f_{46}^\topoid{A446} = \doubleboxb{p_1}{p_2}{q_2}{q_1} \quad  
f_{47}^\topoid{A446} = \doubleboxbtwo{p_1}{p_2}{q_2}{q_1}
\end{eqnarray*}

\begin{eqnarray*}
f_{48}^\topoid{B174} &=& \boxbubbleaB{p_1}{q_1}{q_2}{p_2}  \quad 
f_{49}^\topoid{\thickbar{B}174} = \boxbubbleaB{p_1}{q_2}{q_1}{p_2}  \quad
f_{50}^\topoid{B182} = \boxxaB{p_1}{q_1}{q_2}{p_2}
\end{eqnarray*}

\begin{eqnarray*}
f_{51}^\topoid{B213} &=& \boxxbB{p_1}{q_1}{q_2}{p_2} \quad
f_{52}^\topoid{B213} = \boxxbbbbbB{p_1}{q_1}{q_2}{p_2} \quad
f_{53}^\topoid{B213} = \boxxbbbbbbB{p_1}{q_1}{q_2}{p_2} \\
f_{54}^\topoid{B213} &=& \boxxbbB{p_1}{q_1}{q_2}{p_2} \quad
f_{55}^\topoid{B249} = \boxbubbleOB{p_1}{q_1}{q_2}{p_2} \quad
f_{56}^\topoid{B215} = \boxxbpaB{p_1}{q_1}{q_2}{p_2}\\
f_{57}^\topoid{\thickbar{B}215} &=& \boxxbpaB{p_1}{q_2}{q_1}{p_2}\quad
f_{58}^\topoid{B247} = \doubleboxB{p_1}{q_1}{q_2}{p_2} \quad
f_{59}^\topoid{B247} = \doubleboxBtwo{p_1}{q_1}{q_2}{p_2}
\end{eqnarray*}

\begin{eqnarray*}
 f_{60}^\topoid{C231} = \trianglecrossA{p_{12}}{q_1}{q_2}\quad 
 f_{61}^\topoid{C252} = \trianglecrossB{p_{2}}{p_{23}}{q_1} \quad 
 f_{62}^\topoid{C318} = \trianglecrossC{p_{12}}{p_1}{p_2}
\end{eqnarray*}

\begin{eqnarray*}
 f_{63}^\topoid{C126} &=& \boxxbmcrossxAa{p_2}{q_2}{p_1}{q_1} \quad 
 f_{64}^\topoid{C126} = \boxxbmcrossxAb{p_2}{q_2}{p_1}{q_1} \quad
 f_{65}^\topoid{C207} = \boxxbmcrossxBa{p_1}{p_2}{q_2}{q_1} \\ 
 f_{66}^\topoid{C207} &=& \boxxbmcrossxBb{p_1}{p_2}{q_2}{q_1} \quad
 f_{67}^\topoid{C207} = \boxxbmcrossxBc{p_1}{p_2}{q_2}{q_1} \quad
 f_{68}^\topoid{C207} = \boxxbmcrossxBd{p_1}{p_2}{q_2}{q_1}
\end{eqnarray*}

\begin{eqnarray*}
 f_{69}^\topoid{C239} &=& \doublecrossA{p_1}{p_2}{q_2}{q_1}\quad
 f_{70}^\topoid{C239} = \doublecrossAb{p_1}{p_2}{q_2}{q_1} \quad
 f_{71}^\topoid{C254} = \doublecrossB{p_1}{q_2}{q_1}{p_2}\\
 f_{72}^\topoid{C254} &=& \doublecrossBb{p_1}{q_2}{q_1}{p_2}\quad
 f_{73}^\topoid{C382} = \doublecrossCa{p_1}{p_2}{q_2}{q_1} \quad
 f_{74}^\topoid{C382} = \doublecrossCb{p_1}{p_2}{q_2}{q_1}\\
 f_{75}^\topoid{C382} &=& \doublecrossCc{p_1}{p_2}{q_2}{q_1} \newline
\end{eqnarray*} }

Thick lines denote the massive external particles, dots additional
powers of the corresponding propagator.
We introduced the short-hand notations 
$p_{ij} = p_i+p_j$ and $p_{ijk} = p_i+p_j+p_k$, where $p_3 = -q_1$.
In some cases we denote additional numerators of the integrand,
where the definition of the loop momenta $k$, $l$ is implicitly
given by the corresponding integral family in table~\ref{tab:auxtopo}.
The families $\bar{A}$ and $\bar{B}$ emerge from $A$ and $B$
by swapping the two incoming momenta $p_1$ and $p_2$.

Given any of the integral families $T={A,B,C}$ (and the crossed variants),
every integral is defined with the integration measure:
\begin{equation}
 f_{n}^{T} = \left(\frac{S_\epsilon}{16 \pi^2}\right)^{-2}\, (m^2)^{2 \epsilon }
\int \frac{d^d k}{(2 \pi)^d}\frac{d^d l}{(2 \pi)^d}\,
  \frac{1}{D_{T_1}^{n_1}\,...\,D_{T_9}^{n_9}}\,, \labbel{eq:measure}
\end{equation}
where the indices $T_j$ run over the propagators in the different 
integral families, $d = 4-2\epsilon$ and
\begin{equation}
  S_\epsilon =  (4 \pi)^\epsilon \, {\Gamma(1+ \epsilon) \, 
\Gamma^2(1-\epsilon) \over \Gamma(1-2\epsilon)}\,.
\end{equation}
The transition between $\Re(m^2)>0$ and $\Re(m^2)<0$ is understood
to be taken with $\Im(m^2)>0$.

Let us note here that, as already discussed above, 
the main point of our bottom-up construction of the canonical basis 
is that we do not need to look at the global properties of the
$75 \times 75$ matrix, but we can move
step by step for increasing values of $t$, treating separately 
the differential equations for different sectors 
which are topologically disentangled from each other. 
In this sense we need to verify \textsl{separately for every topology} 
that the two requirements of $(1)$ triangularity as $\epsilon \to 0$,
and $(2)$ linearity in $\epsilon$, are satisfied \textsl{\underline{only} for the 
homogeneous part of the sub-system}\footnote{Note once more that,
while this is enough to ensure that the property of triangularity
is true for the whole system, the same is in general not true
for the linearity in $\epsilon$.}.\newline

Deriving the differential equations for the basis above one 
immediately sees that:

\begin{enumerate}
 \item All sub-systems of differential equations, topology by topology,
       have the property that their homogeneous part is triangular 
       (or even decouples) in the limit $\epsilon \to 0$.
       
 \item On the other hand \textsl{almost} all sub-systems fulfil the 
 property of linearity in $\epsilon$, except six, which are:
$$\{f_{18}^\topoid{A166},f_{19}^\topoid{A166}\},\quad
 \{f_{20}^\topoid{A198},f_{21}^\topoid{A198}\},\quad
 \{f_{29}^\topoid{A181},f_{30}^\topoid{A181}\},$$
$$ \{f_{31}^\topoid{\thickbar{A}181},f_{32}^\topoid{\thickbar{A}181}\},\quad
 \{f_{33}^\topoid{A182},f_{34}^\topoid{A182}\},\quad
 \{f_{35}^\topoid{\thickbar{A}182},f_{36}^\topoid{\thickbar{A}182}\}.$$
 In particular the homogeneous systems contain
 terms proportional to $\epsilon^2$ and $1/(1-2\epsilon)$.
\end{enumerate}

In order to put these sub-systems in the right form (note that four of them are
equal two-by-two under a permutation of the external momenta) one can proceed
in different ways. Since the systems are very simple (in all cases $2\times2$ systems)
one can study directly the homogeneous parts of the latter and, one-by-one,
look for appropriate linear combinations of the two MIs which
remove the terms in $\epsilon^2$  and $1/(1-2\epsilon)$. 
In particular in the case of
sectors $181$ and $182$ (and their crossing) it is easy to see that
these unwanted terms are removed by a simple rescaling 
of the two masters.

For the first two sectors ($A166$ and $A198$),
this is not enough and a proper linear combination of
the two MIs in the sector is needed.
Nevertheless, having the exact solution for all planar MIs 
at hand~\cite{Gehrmann:2013cxs} it is clear that, 
for both sectors, the second MI (the one with a dotted propagator 
$f_{19}^\topoid{A166}$, $f_{21}^\topoid{A198}$) 
is already in the right form (i.e. it is a function of uniform transcendentality
multiplied by a single rational factor), and therefore must not be changed.
This implies that only the first MI must be substituted by a linear
combination of the two, which can be easily found imposing that also
the latter becomes a function of uniform transcendentality multiplied
by a single rational factor (or, equivalently, that the homogeneous
part of the system contains only a linear dependence on $\epsilon$).

Once the differential equations for these four sectors have been put in the right
form we can easily apply our algorithm to all remaining MIs.
As a result we get a new basis $\vm = \{m_j\}$ with $j=1,...,75$ whose differential
equations with respect to both external invariants $(x,z)$ are in canonical form.
We enclose the explicit definition of the basis 
expressed in terms of the initial choice of MIs depicted above in Appendix~\ref{app:basis}.

\subsection{Comments on the basis change}

Following the construction of the canonical basis as
 described in the previous section, it appears clear how, at least in the case
 under consideration, the issue of finding a 
 canonical basis for a set of master integrals can be identified to that of
 being able to integrate out the homongeneous part of the system in $\epsilon=0$
 in terms of rational functions only\footnote{The same seems anyway to apply
 also to more involved cases where a preliminary analysis has already been
 carried out. Moreover note that similar conclusions on the behaviour
 of the differential equations for $\epsilon=0$ are drawn in~\cite{Caron-Huot:2014lda}.}. 
 In this sense, having a basis of MIs whose differential equations 
 are in canonical form is, for \textsl{practical purposes}, 
 almost equivalent to having a basis whose differential equations
 are \textbf{triangular} for $\epsilon=0$, and whose homogeneous
 parts can be integrated in terms of rational functions only.
 From this point of view, casting the system 
 of differential equations into canonical form consists 
 in separating into two different steps
 the two conceptually different issues of: $(a)$ integrating the homogeneous 
 parts of the equations (which provide the somehow ``\textsl{trivial}'' 
 rational prefactors of the MIs), and $(b)$ integrating the ``\textsl{non-trivial}''
 dependence of the master integrals on \textsl{transcendental functions} (and in
 particular on GHPLs). It looks reasonable to think that such 
 a ``factorization'' can be achievable as long as the master integrals 
 are expressed in terms of GHPLs only (and Chen Iterated
 integrals in general). On the other hand, it is not clear how and if this structure
 could be preserved or generalized in the case of more complicated
 functional behaviours.
 
It is also important to stress that, both if the equations
  are in canonical form, and if the equations are triangular for $\epsilon=0$
  \textsl{(with the homogeneous part solvable in terms of rational functions
  only)}, the integration of the differential equations becomes straight-forward.
  A task which remains is the determination of the boundary terms,
  which is non-trivial in particular in the case of non-planar integrals.
  In this regard, having the equations in canonical form
  does not solve any conceptual difficulties by itself.
In either case, the canonical basis gives not only a clearer view on the
structure of the problem, it also has great practical advantages due to
the simpler and more compact expressions which need to be handled in this
approach.

\subsection{Differential equations}

Given the basis in Appendix~\ref{app:basis} we can derive differential equations
in both independent variables $(x,z)$.
As already anticipated the equations take the canonical form~\eqref{eq:cform}:
\begin{align}
d\,\vm(\epsilon;x,z) &= 
 \epsilon\, d A(x,z)\,\vm(\epsilon;x,z) \labbel{eq:deqqqbvv}
\end{align}
where the differential $d$ acts on the two variables $x,z$.
The matrix $A(x,z)$ does not depend on $\epsilon$ and it can be decomposed as
\begin{equation}
 A(x,z) = \sum_{k=1}^{10}\, A_k\, \ln{(r_k)}\,,
\end{equation}
where the $A_k$ are \textsl{constant matrices}, whose entries are in particular
just \textsl{rational numbers}, and the $r_k$ are polynomial
functions of $(x,z)$ and constitute the 
so-called alphabet of the GHPLs which will be needed in order to integrate the equations:
\begin{align}
 r_k = &\left\{\, x,\,1-x,\,1+x,\,z,\,1+z,\,x-z,\,1-xz, \right.\nonumber 
 \\& \left. \,1+x^2-xz,\,1+x+x^2-xz,\,z(1+x+x^2)-x\,\right\}\,.\labbel{eq:alphabet}
\end{align}

Expanding in $\epsilon$, the canonical form
ensures full decoupling  of the differential equations~\eqref{eq:deqqqbvv}
order by order in $\epsilon$.
The integration \textsl{up to a boundary term} becomes trivial
and can be carried out entirely algebraically.

\section{Integration and boundary conditions}
\labbel{sec:boundaries}

We consider the full system of differential equations for all 75 master
integrals in a uniform manner.
Our normalization is such that the solutions for our master integrals
have a Taylor expansion,
\begin{equation}
\vm(\epsilon;x,z) = \sum_{i=0}^{\infty} \vm^{(i)}(x,z) \epsilon^i\,, \labbel{eq:laurentm}
\end{equation}
where the weight 0 contributions start at $\epsilon^0$.
We solve the full vector of coefficient functions $\vm^{(i)}$ order by order
in $\epsilon$ up to and including weight 4.

For master integrals depending on $z$ we choose to integrate the partial
differential equation in $z$ for fixed $x$ implied by~\eqref{eq:deqqqbvv}.
This gives us the solution up to a function of $x$, which needs to be
fixed by additional constraints discussed below.
For master integrals independent on $z$ we integrate the partial differential
equation in $x$, which determines the solution up to a constant.
It is obvious that this procedure naturally leads to iterated integrals.
The $d\ln$ form of the differential equations ensures that the
iterated integrals can be expressed in terms of Goncharov's multiple
polylogarithms
\begin{align}
G(w_1, w_2, \cdots, w_n; z) &\equiv \int_0^z \mathrm{d}t \, 
\frac{1}{t - w_1} G(w_2, \cdots, w_n; t) \, ,\\
G( \underbrace{0,\cdots, 0}_n; z) &\equiv \frac{1}{n!} \ln^n z\,.
\end{align}
Here, the $w_i$ are complex rational functions of the indeterminants.
To handle non-linear letters we also employ
generalized weights~\cite{vonManteuffel:2013vja}.
\begin{equation}
G( [f(o)], w_2, \cdots, w_n; z)
  = \int_0^z dt \, \frac{f'(t)}{f(t)} G(w_2, \cdots, w_n; t) \, ,
\end{equation}
where $f(o)$ is an irreducible rational polynomial and $o$ is a dummy variable.

In order to fix the boundary terms we use two ingredients.
For some of the simplest integrals, namely a small number of tadpole, bubble and triangle integrals,
we use their known analytic solutions from the
literature~\cite{Gehrmann:2013cxs,vanNeerven:1985xr}.
For all other integrals we require the absence of logarithmic divergencies
for the solutions in certain kinematical limits.
This requires the linear combinations of master integrals multiplying the corresponding $d\ln$
terms in the differential equations to vanish in the respective limit.
This completely fixes the remaining boundary terms, i.e.\ the unknown functions of $x$ respectively the
constants.

Since we consider also non-planar integrals it is unavoidable to deal with cuts in
$s$, $t$ and $u$ at the same time, i.e.\ we have to handle uncrossed and
crossed kinematics.
From the Feynman parameter representation it is clear that there is
a Euclidean region with $s$, $t$, $u$, $m^2$ all less than zero, such that the integrals
are real.
From the on-shell relation~\eqref{eq:xyz} one sees, however, that it is not possible
to parametrize this region employing real valued parameters $x$, $z$ and $m^2$.
Note that in the scheme~\cite{vonManteuffel:2013uoa} employed here,
the solutions develop explicit and implicit imaginary parts already during the
iterative integration procedure.

We require regularity of each integral in some of the following collinear and,
depending on its cut structure, threshold limits:
\begin{equation}
z \to x, \qquad
z \to 1/x,\qquad
z \to -1,\qquad
z \to (1+x+x^2)/x,\qquad
x \to 1\,.
\end{equation}
We emphasize that we impose these conditions for points in the unphysical
region, the algebraically equivalent limits in the physical region may
actually be divergent due to branch cuts.
The difference between the two cases lies in the way the signs of the 
imaginary parts of the parameters needs to be chosen when approaching
the respective point, as dictated by the Feynman propagator $i0$
prescription.

We assign a small positive imaginary part to $s$, $t$, $u$ and $m^2$ to 
fix branch cut ambiguities.
While the $m^2$ dependence is not explicite in our dimensionless master
integrals $\vm(\epsilon;x,z)$ anymore, we anticipate its former presence
with $\Re(m^2)<0$.
This translates to (small) imaginary parts $\Im(x) >0$, $\Im(z) < 0$
and $\Im(y) = \Im ((1 + x^2 - x z)/x) > 0$.
The limits were computed using in-house Mathematica packages for
multiple polylogarithms~\cite{andreasgpl,erichgpl},
where we employed both, coproduct based and non-coproduct based
limit algorithms.
In practice we employ small but finite imaginary parts, such that the
complex parameters fulfil the on-shell relation~\eqref{eq:xyz}.
We match the constants appearing in the limit computations by a numerical
fitting procedure.
This step utilizes the numerical evaluation of multiple polylogarithms
\cite{Vollinga:2004sn} in GiNaC \cite{Bauer:2000cp}.

\section{Solutions and checks}
\labbel{sec:checks}

We obtain the solutions in terms of GHPLs of argument
$z$ and weights $\{0, -1, x, 1/x, (1+x^2)/x, x/(1+x+x^2),
(1+x+x^2)/x\}$ and GHPLs of argument $x$ and weights $\{0, -1, 1,
[1+o^2], [1+o+o^2]\}$.
The explicit expressions are rather lengthy and therefore provided
via an ancillary file on the arXiv only.

We performed several checks on the results.
First of all, we integrated the whole $75 \times 75$ system of 
differential equations at once, fixing consistently all boundary
conditions using the limits described above.
We explicitly verified that our solutions fulfil the partial differential
equations both in $x$ and in $z$.
This is a non-trivial check for integrals depending on both variables,
for which we fixed $x$-dependent boundary terms by regularity conditions.

As a subset of the integrals considered here,
we re-calculated all non-trivial planar master integrals presented
in~\cite{Gehrmann:2013cxs}.
Taking $z' \to z'+i0$ and swapping masters 3 and 4 of sector B213
in the result of that reference, we translate these expressions
to our new functional basis and find perfect agreement at the
analytical level.

For the previously unknown non-planar master integrals we compared
our results against numerical samples obtained with the sector
decomposition program SecDec2~\cite{Borowka:2012yc,Borowka:2013cma}.
We found the program particularly useful since it allowed us to perform checks
of our results both in the Euclidean and in the physical region.
In the Euclidean region we set $x$ to a truly complex number.
Note that to obtain a real number at all consists already in 
a very non-trivial check of our solution.
In particular, 
we could verify all our master integrals in the Euclidean region
with a typical precision of at least 4 digits for the weight 4
coefficients. On the other hand, the numerical evaluation by
sector decomposition in the Minkowski region was much more cumbersome.
Using SecDec2 we could evaluate all \textsl{corner integrals}
(integrals with no dots nor scalar products) up to weight 4,
finding good agreement with our result. For sectors
involving more than one master integral we additionally considered integrals
with dots and/or scalar products. For them we could check at least the
weight 3 contributions, in some cases also the weight 4 parts.
The combination of these checks in the Euclidean and in the Minkowski
region provides stringent evidence for the correctness of our results.

\section{Real valued functions and expansions}
\labbel{sec:realfuncs}

For the purpose of numerical evaluation in the physical region
the primary form of our solutions is not optimal yet, e.g.\ because
the multiple polylogarithms are not single valued and their numerical
evaluation is not straightforward.
We follow the procedure described in~\cite{Duhr:2011zq} and project onto
a new functional basis which consists of $\Li_{2,2}$,
classical polylogarithms $\Li_n$ $(n=2,3,4)$ and logarithms.
The $\Li$-functions are related to the $G$-functions via
\begin{equation}
\Li_n (x_1) = -G(\underbrace{0,\cdots,0,1}_{n};x_1)  \,, \qquad 
\Li_{2,2}(x_1,x_2) = G\left(0, \frac{1}{x_1}, 0, \frac{1}{x_1 x_2}; 1\right) \,.
\end{equation}
In our new functional basis we allow for rather complicated rational
functions of $x$ and $z$.
We choose them such that the functions are real valued and the imaginary
parts of the solutions are explicit over the entire physical domain.

In~\cite{Bonciani:2013ywa} it was demonstrated that this method works
also in the presence of generalized weights, which could in fact
be eliminated at the level of the amplitude.
In the present case we work at the level of the master integrals.
Also here, we successfully apply this projection onto real valued
functions and eliminate all generalized weights $\{[1+o^2], [1+o+o^2]\}$
using a coproduct based algorithm.

We can actually go one step further and restrict the target function
space even more. For the functions $\Li_n(x_1)$, $\Li_{2,2}(x_1,x_2)$
we select real arguments with
\begin{equation}
|x_1| < 1\,, \qquad |x_1 x_2| < 1\,.
\end{equation}
In this way, the multiple polylogarithms are not only real valued but
correspond directly to a convergent power series expansion
\begin{align}
\Li_n(x_1) &= -\sum_{j_1 = 1}^{\infty} \frac{x_1^{j_1}}{j_1^n},\\
\Li_{2,2}(x_1,x_2) &= \sum_{j_1 = 1}^{\infty}\sum_{j_2 = 1}^{\infty}
  \frac{x_1^{j_1}}{(j_1+j_2)^2} \frac{(x_1 x_2)^{j_2}}{j_2^2}
\end{align}
see e.g.\ eq.\ (20) of \cite{Vollinga:2004sn}.
While it is not a priori obvious that such a restricted set of functions
is sufficient to represent our master integrals, we find that this is indeed
the case.
Our choice of functions drastically improves the numerical evaluation time,
since it avoids additional transformations which would be required otherwise
to map to an appropriate expansion.
Evaluating all master integrals discussed in this paper takes only
fractions of a second in a generic phase space point on a single core.

For completeness, we also expand our solutions both at the
production threshold and in the small mass region.
The threshold region is characterized by $\beta \to 0$ for fixed
$\cos\theta$, where $\beta=\sqrt{1-4m^2/s}$ is the velocity of each vector boson
and $\theta$ the scattering angle in the center-of-mass frame, such that
$z = 1 + 2 \beta (\beta+\cos\theta)/(1-\beta^2)$.
We find it convenient to directly expand our full solutions in the
real-valued function representation, rather than the individual $G$-functions.
The expansion contains $\beta$, $\ln\beta$, GHPLs of argument $\cos\theta$
and weights $\{-1, 1\}$ as well as the constants
$\ln(2)$ and $\Li_4(1/2)$.
Similarly, we consider the small mass limit $m^2/s \to 0$ for fixed
$\phi=-(t-m^2)/s$.
The expansion contains $m^2/s$, $\ln(m^2/s)$ and GHPLs of argument
$\phi$ and weights $\{0,1\}$.
The first couple of orders for both expansions as well as our results
in terms of real-valued functions are provided via ancillary files on the arXiv.

\section{Conclusions}
\labbel{sec:conc}

In this paper, we computed the full set of master integrals relevant to the 
two-loop QCD corrections to the production of 
two vector bosons of equal mass in the collision of massless partons.
These two-loop four-point functions 
are computed using the differential equation 
method~\cite{Kotikov:1990kg,Remiddi:1997ny,Caffo:1998du,Gehrmann:1999as}.
 We describe in detail how we find a  canonical 
basis~\cite{Henn:2013pwa} for the master integrals. 
In this basis, the differential equations for the master integrals can be solved in an elegant 
and compact manner in terms of iterated integrals.
These general solutions are then matched onto appropriate boundary values,
requiring non-trivial transformations of the iterated integrals.
Our analytical results for all master integrals are expressed in terms of multiple polylogarithms,
they are provided with the arXiv submission of this article.
We find that it is possible to employ a restricted set of multiple polylogarithms,
which allows for a particularly fast and precise numerical evaluation.
We validated our solutions against numerical samples obtained using
sector decomposition.

With the full set of master integrals 
derived in this paper, it is now possible to derive the two-loop 
corrections to the amplitudes for 
$q\bar q\to W^+W^-$ and $q\bar q \to ZZ$, and to compute the 
NNLO corrections to the pair production of massive vector bosons. Combined with 
precision measurements of these observables at the LHC, these results will allow for 
a multitude of tests of the electroweak theory at unprecedented precision.

\section*{Acknowledgements}

We are grateful to Sophia Borowka and Gudrun Heinrich for their assistance 
with SecDec2 and to Pierpaolo Mastrolia for interesting comments
on the manuscript. 
AvM would like to thank Stefan Weinzierl for solving issues with
the GiNaC implementation of the multiple polylogarithms and
Andrea Ferroglia for useful discussions.
We acknowledge interesting discussions with 
Johannes Henn and Pierpaolo Mastrolia on the properties of the canonical basis.
Finally we thank Kirill Melnikov for comparison of numerical
results from~\cite{Caola:2014lpa} prior to publication.
This research was supported in part by
the Swiss National Science Foundation (SNF) under contract
PDFMP2-135101 and  200020-149517, as well as  by the European Commission through the 
``LHCPhenoNet" Initial Training Network PITN-GA-2010-264564 and the ERC Advanced Grant
``MC@NNLO" (340983).
The work of AvM\ was supported in part by the Research Center {\em Elementary
Forces and Mathematical Foundations (EMG)} of the Johannes Gutenberg University
of Mainz and by the German Research Foundation (DFG).

\appendix

\section{Canonical basis}
\labbel{app:basis}
As a result of the algorithm described in section~\ref{sec:basis} 
we find the following canonical basis, which for simplicity 
is also attached to the arXiv submission of this paper:

{\footnotesize

\begin{align*}
 &m_1 =   \epsilon^2 \, \frac{m^2 (1+x)^2}{x}\,f_{1}^\topoid{A38}\,,\quad 
 m_2 =    \epsilon^2 \,        m^2 f_{2}^\topoid{A134}\,,\quad
 m_3 =    \epsilon^2 \,        m^2 z \,f_{3}^\topoid{A148}\,,\quad
 m_4 =    \epsilon^2 \,  \frac{m^2(1+x^2-xz)}{x}f_{4}^\topoid{\thickbar{A}148}\,,\\
 &m_5 =   \epsilon^2 \,  \frac{m^4(1+x)^4}{x^2} f_{5}^\topoid{A99}\,,\quad 
 m_6 =    \epsilon^2 \,  \frac{m^4(1+x)^2}{x}  f_{6}^\topoid{A195}\,,\quad 
 m_7 =    \epsilon^2 \,        m^4 f_{7}^\topoid{A387}\,,\quad 
 m_8 =   -\epsilon^2 \,        m^4 z\, f_{8}^\topoid{A394}\,,\\
 &m_9 =  -\epsilon^2 \,  \frac{m^4(1+x^2-xz)}{x}f_{9}^\topoid{\thickbar{A}394}\,,\quad 
 m_{10} = \epsilon^2 \,        m^4 z^2 f_{10}^\topoid{A408}\,,\quad
 m_{11} = \epsilon^2 \,  \frac{m^4(1+x^2-xz)^2}{x^2} f_{11}^\topoid{\thickbar{A}408}\,,\\
 &m_{12}= \epsilon^2 \,        m^4 f_{12}^\topoid{A418}\,,\quad
 m_{13}= -\epsilon^3 \,  \frac{m^2(1+x)^2}{x}f_{13}^\topoid{A53}\,,\quad 
 m_{14}=  \epsilon^3 \,        m^2(1+z) f_{14}^\topoid{A142}\,,\\
 &m_{15}= \epsilon^3 \,  \frac{m^2(1+x+x^2-xz)}{x}f_{15}^\topoid{\thickbar{A}142}\,,\quad 
 m_{16} =-\epsilon^3 \,        m^2(1+z) f_{16}^\topoid{A149}\,,\quad
 m_{17} =-\epsilon^3 \,  \frac{m^2(1+x+x^2-xz)}{x}f_{17}^\topoid{\thickbar{A}149}\,,\\
 &m_{18}= \epsilon^3 \,  \frac{m^2(1-x^2)}{x}f_{19}^\topoid{A166}\,,\quad
 m_{19} = \epsilon^2 \left[ (1-2\epsilon)(1-3\epsilon)\,f_{18}^\topoid{A166}
                            + \frac{m^2(1+x)^2}{2\,x}\,f_{1}^\topoid{A38} 
                           \right]
         -\epsilon^3 \,  \frac{m^2(1+x)}{x}f_{19}^\topoid{A166}\,,\\
 &m_{20} = \epsilon^3 \,  \frac{m^2(1-x^2)}{x}f_{21}^\topoid{A198}\,,\quad   
  m_{21} = \epsilon^2 \left[ (1-2\epsilon)(1-3\epsilon)\,f_{20}^\topoid{A198}
                            - m^2\,f_{2}^\topoid{A134} 
                           \right]
         -\epsilon^3 \,  \frac{m^2(1-2\,x^2)}{x}f_{21}^\topoid{A198}\,,\\
 &m_{22} = \epsilon^3\,   \frac{m^4(1-x)(1+x)^3}{x^2}\, f_{22}^\topoid{A227}\,,\quad
  m_{23} = \epsilon^3 \,  \frac{m^4(1-x^2)}{x}f_{23}^\topoid{A419}\,,\quad 
  m_{24} = \epsilon^4 \,  \frac{m^2(1-x^2)}{x}f_{24}^\topoid{A199}\,,\\
 &m_{25} = \epsilon^4 \,       {m^2(1+z)}f_{25}^\topoid{A398}\,, \quad
  m_{26} = \epsilon^4 \,  \frac{m^2(1+x+x^2-xz)}{x}f_{26}^\topoid{\thickbar{A}398}\,,\quad
  m_{27} = \epsilon^4 \,  \frac{m^2(1-x^2)}{x}f_{27}^\topoid{A422}\,,\\
 &m_{28} = \epsilon^3 \,  \frac{m^4z\,(1+x)^2}{x} f_{28}^\topoid{A174}\,,\quad
  m_{29} = \epsilon^3 (1-2\epsilon)\,\frac{m^2(1-x^2)}{x} f_{29}^\topoid{A181}\,\,\quad
  m_{30} = \epsilon^3 \,  \frac{m^4z\,(1+x)^2}{x} f_{30}^\topoid{A181}\,,\\
 &m_{31} = \epsilon^3 (1-2\epsilon)\,\frac{m^2(1-x^2)}{x} f_{31}^\topoid{\thickbar{A}181}\,\,\quad
  m_{32} = \epsilon^3 \,  \frac{m^4(1+x)^2(1+x^2-xz)}{x^2} f_{32}^\topoid{\thickbar{A}181}\,,\\
 &m_{33} = \epsilon^4 \,  \frac{m^2(1+x+x^2-xz)}{x} f_{33}^\topoid{A182}\,,\quad
  m_{34} = \epsilon^3 \,       {m^4(1+z)}  f_{34}^\topoid{A182}\,,\quad
  m_{35} = \epsilon^4 \,       {m^2(1+z)} f_{35}^\topoid{\thickbar{A}182}\,,\\
 &m_{36} = \epsilon^3 \,  \frac{m^4(1+x+x^2-xz)}{x} f_{36}^\topoid{\thickbar{A}182}\,,\quad
  m_{37} = \epsilon^3 \,  \frac{m^4z\,(1+x)^2}{x}f_{37}^\topoid{A214}\,, \quad
  m_{38} = \epsilon^3 \,  \frac{m^4 (1+x)^2(1+x^2-xz)}{x^2}f_{38}^\topoid{\thickbar{A}214}\,, \\
 &m_{39} = \epsilon^3 \,  \frac{m^6z\,(1+x)^2}{x}f_{39}^\topoid{A427}\,, \quad
  m_{40} = \epsilon^4 \,  \frac{m^4(1+z)(1+x)^2}{x}f_{40}^\topoid{A215}\,,\quad
  m_{41} = \epsilon^4 \,  \frac{m^4(1+x+x^2-xz)(1+x)^2}{x^2}f_{41}^\topoid{\thickbar{A}215}\,,\\
 &m_{42} = \epsilon^4 \,  \frac{m^4(z(1+x+x^2)-x)}{x}f_{42}^\topoid{A430}\,,\quad
  m_{43} = \epsilon^4 \,  \frac{m^6z\,(1+x)^4}{x^2}f_{43}^\topoid{A247}\,, 
  \end{align*}
  \begin{align*}
   m_{44} = &- \epsilon^2 \frac{m^2 (1 + x)^2 }{2\,x z}f_{1}^\topoid{A38}
  + \epsilon^2  \frac{5m^2}{2\,z} f_{2}^\topoid{A134}  
  + \epsilon^2  \frac{9m^2}{2} f_{3}^\topoid{A148}   
  + \epsilon^3  \frac{6m^2(1 + z)}{z}   f_{16}^\topoid{A149}
  + \epsilon^2 \frac{4(1 - 2 \epsilon) (1 - 3 \epsilon)}{z}   f_{18}^\topoid{A166}\\
  &- \epsilon^3  \frac{2m^2(1 + x)^2 }{x z} f_{19}^\topoid{A166}
  + \epsilon^3 \frac{2m^4 (1 + x)^2 }{x} f_{30}^\topoid{A181}
  + \epsilon^4  \frac{6m^2(1 + x + x^2 - x z) }{x z}f_{33}^\topoid{A182} 
  + \epsilon^3 \frac{4 m^4 (1 + z) }{z} f_{34}^\topoid{A182}\\
  &+ \epsilon^4 \frac{ m^4(1 + x)^4 }{x^2} f_{44}^\topoid{A247}\,,
  \end{align*}
  \begin{align*}
 &m_{45} = \epsilon^4   \frac{m^4(1-x)(1+x)^3}{x^2}f_{45}^\topoid{A247}\,,\quad
  m_{46} = \epsilon^4   \frac{m^6z^2(1+x)^2}{x}f_{46}^\topoid{A446}\,,\\
  &m_{47} = \epsilon^4 m^4 z\,\left(  \frac{\,(1+x)^2}{x}f_{42}^\topoid{A430}
         +        {\,(1+z)} f_{47}^\topoid{A446} \right)\,,
  m_{48} = \epsilon^3 \,  \frac{m^4(x-z)(1-xz)}{x}f_{48}^\topoid{B174}\,,\\
 &m_{49} = \epsilon^3 \,  \frac{m^4(x-z)(1-xz)}{x}f_{49}^\topoid{\thickbar{B}174}\,,\quad
  m_{50} = \epsilon^4 \,  \frac{m^2(1+x)^2}{x}f_{50}^\topoid{B182}\,,\quad
  m_{51} = \epsilon^4 \,  \frac{m^2(1-x^2)}{x}f_{51}^\topoid{B213}\,,\\
 &m_{52} = \epsilon^3 \,  \frac{m^4(1+x+x^2-xz)}{x}f_{52}^\topoid{B213}\,,\quad
  m_{53} = \epsilon^3 \,         {m^4(1+z)} f_{53}^\topoid{B213}\,,\quad
  m_{54} = \epsilon^3 \,  \frac{m^4(x-z)(1-xz)}{x}f_{54}^\topoid{B213}\,,\\
  &m_{55} = \epsilon^3 \,\frac{m^6(x-z)(1-xz)}{x}f_{55}^\topoid{B249}\,,\quad
  m_{56} = \epsilon^4 \,\frac{m^4(1+z)(1+x^2-xz)}{x}f_{56}^\topoid{B215}\,,\\
 &m_{57} = \epsilon^4 \,\frac{m^4z\,(1+x+x^2-xz)}{x}f_{57}^\topoid{\thickbar{B}215}\,,\quad
 m_{58} = \epsilon^4 \,\frac{m^6(x-z)(1-xz)(1+x^2-xz)}{x^2}f_{58}^\topoid{B247}\,,
  \end{align*}
\begin{align*}
 m_{59} =  &-\epsilon^2\, \frac{3m^2 (1+x+x^2-xz)}{2(x-z)(1-xz)}f_{2}^\topoid{A134}
      -\epsilon^2\, \frac{3m^2 z\,(1+x+x^2-xz)}{4(x-z)(1-xz)} f_{3}^\topoid{A148}\\
    &-\epsilon^2\,  \frac{3m^2(1+x^2-xz) (1+x+x^2-xz)}{4x(x-z)(1-xz)}f_{4}^\topoid{\thickbar{A}148}
      -\epsilon^3\, \frac{m^4(1+x+x^2-xz)}{x}f_{49}^\topoid{\thickbar{B}174}\\
     &+\epsilon^4\,\frac{3 m^2 (1+x)^2(1+x+x^2-xz)}{x(x-z)(1-xz)}f_{50}^\topoid{B182}
      +\epsilon^4\,\frac{m^4 (1+x^2-xz)(1+x+x^2-xz)}{x^2}f_{59}^\topoid{B247}\,,
\end{align*}
\begin{align*}
 &m_{60} = \epsilon^4   \frac{m^4(1-x^2)^2}{x^2}f_{60}^\topoid{C231}\,,\quad
  m_{61} = \epsilon^4        {m^4(1+z)^2}f_{61}^\topoid{C252}\,,\quad
  m_{62} = \epsilon^4   \frac{m^4(1+x)^4}{x^2}f_{62}^\topoid{C318}\,,\\
 &m_{63} = \epsilon^4   \frac{m^4(1+x)^2}{x}f_{63}^\topoid{C126}\,,\quad
  m_{64} = \epsilon^4   {m^2(1+z)}\left[ f_{64}^\topoid{C126} - f_{33}^\topoid{A182} \right]\,,\quad
  m_{65} = \epsilon^4   \frac{m^4(1+x)^2}{x}f_{65}^\topoid{C207}\,,
\end{align*}
\begin{align*}
 m_{66}  = &+\epsilon^2 \frac{m^2(1+x)^2}{4x(1+z)} \left[f_{2}^\topoid{A134} + z\,f_{3}^\topoid{A148} \right]
         + \epsilon^3 \frac{m^2(1+x)^2}{x}f_{16}^\topoid{A149}
         - \epsilon^4 \frac{m^4(1+x)^2(1+x^2-xz)}{x^2}f_{65}^\topoid{C207}\\
         &+\epsilon^4 \frac{m^2(1+x+x^2-xz)}{x}f_{66}^\topoid{C207}\,,\\
 m_{67}  = &+\epsilon^2 \frac{m^2(1+x)^2}{4(1+x+x^2-xz)} \left[f_{2}^\topoid{A134} 
           + \frac{(1+x^2-xz)}{x}\,f_{4}^\topoid{\thickbar{A}148} \right]
         + \epsilon^3 \frac{m^2(1+x)^2}{x}\left[ f_{17}^\topoid{\thickbar{A}149}
         + \epsilon\, f_{33}^\topoid{A182} \right] \\
         &-\epsilon^4 {m^2(1+z)}\left[ f_{51}^\topoid{B213} + f_{66}^\topoid{C207} - f_{67}^\topoid{C207} \right]\,,\\
 m_{68}  = &+ \epsilon^4 \frac{m^2(1-x^2)}{x} \left[ f_{68}^\topoid{C207} - f_{33}^\topoid{A182} \right]\,,\\
 m_{69} =
       &+ \epsilon^4\, \frac{m^4 (1-x^2)(1 + z) }{x} f_{65}^\topoid{C207}
       - \epsilon^4\, \frac{m^4 (1-x^2) (1 + x)}{x^2} f_{60}^\topoid{C231}
       - \epsilon^4\, \frac{m^4 (1-x^2) (1 + x)^2}{x^2}   \left[ f_{41}^\topoid{\thickbar{A}215} - f_{70}^\topoid{C239} \right]\,,\\
 m_{70} = &+ \epsilon^4\, {m^4 (1 + x) (1 + z)} f_{65}^\topoid{C207}
       + \epsilon^4\, \frac{m^4 (1 + x)^2 (1-xz)}{2x^2}   \left[ f_{60}^\topoid{C231} - 2\,f_{70}^\topoid{C239} \right]\\
       &+ \epsilon^4\, \frac{m^6 (1 + x)^2 (1-xz) (x-z)}{2 x^2} f_{69}^\topoid{C239}
       - \epsilon^4\, \frac{m^4 (1 + x)^3}{x} f_{41}^\topoid{\thickbar{A}215}\,,\\
 m_{71} = &+ \epsilon^4\, {m^4 (1 + z)} \left[ f_{42}^\topoid{A430} + z f_{72}^\topoid{C254} \right]
       + \epsilon^2\, \frac{3m^2 z\,(1 + x)^2}{2(1-xz)(x-z)} f_{2}^\topoid{A134}
       + \epsilon^2\, \frac{3m^2 z^2 (1 + x)^2}{4(1-xz) (x-z)} f_{3}^\topoid{A148}\\
       &+ \epsilon^3\, \frac{m^4 z\,(1 + x)^2}{x} \left[ f_{48}^\topoid{B174} + \epsilon f_{65}^\topoid{C207} \right]
       + \epsilon^2\, \frac{3 m^2 z\,(1 + x)^2 (1+x^2-xz)}{4(1-xz) (x-z) x} f_{4}^\topoid{\thickbar{A}148}
       - \epsilon^4\, \frac{3 m^2 z\,(1 + x)^4}{x\,(1-xz) (x-z) } f_{50}^\topoid{B182}\,,\\
m_{72} = &- \epsilon^4\, \frac{m^4 (1-xz)(x-z)}{x\,(1 + z)} \left[ f_{42}^\topoid{A430} + z\,f_{72}^\topoid{C254} \right]
       - \epsilon^3\, \frac{2 m^2 z\, (1 + x)^2}{(1 + z) x} \left[
          f_{14}^\topoid{A142}
          + f_{16}^\topoid{A149}
          - f_{17}^\topoid{\thickbar{A}149}
          \right]\\
       &- \epsilon^4\, \frac{2 m^2 z (1 + x)^2}{(1 + z) x}  \left[
           f_{25}^\topoid{A398}
          - f_{35}^\topoid{\thickbar{A}182}
          + f_{51}^\topoid{B213}
          - f_{64}^\topoid{C126}
          + f_{66}^\topoid{C207}
          - f_{67}^\topoid{C207}
          \right]
       - \epsilon^2\, \frac{3 m^2 z^2 (1 + x)^2}{2(1 + z)^2 x} f_{3}^\topoid{A148}\\
      &- \epsilon^3\, \frac{m^4 z\,(1 + x)^2}{(1 + z) x} \left[ f_{34}^\topoid{A182} - f_{53}^\topoid{B213} \right]
       - \epsilon^4\, \frac{m^4 z\,(1 + x)^2 }{2x} f_{61}^\topoid{C252}
       - \epsilon^2\, \frac{m^2 z\,(1 + x)^2 (3+2x-4xz+3x^2)}{2(1 + z)^2 (1+x+x^2-xz) x} f_{2}^\topoid{A134}\\
      &+ \epsilon^4\, \frac{m^6 z\,(1 + x)^2 (1-xz)(x-z)}{2(1+z) x^2}f_{71}^\topoid{C254}
       + \epsilon^2\, \frac{m^2 z\,(1 + x)^2 (1+x^2-x z)}{2x\,(1 + z) (1+x+x^2-xz)} f_{4}^\topoid{\thickbar{A}148}\\
      &+ \epsilon^4\, \frac{m^4 (1 + x)^2 ( - z + 3 x + 2 x z + x z^2 - x^2 z)}{2(1 + z) x^2} 
                      \left[ f_{63}^\topoid{C126}- f_{65}^\topoid{C207} \right]\,,
       \end{align*} 
       
\begin{align*}
    m_{73} =
       &- \epsilon^4\, \frac{m^4 (1 + x)^2 z}{x}  \left[ f_{63}^\topoid{C126} - f_{74}^\topoid{C382} \right]
       + \epsilon^2\, \frac{m^2 (1 + x)^2 (1 + x^2)}{4 (1+x^2-xz) x}
         \left[ f_{1}^\topoid{A38} + 4\, \epsilon\,f_{19}^\topoid{A166} \right]
       - \epsilon^3\, \frac{m^4 (1 + x)^2 (1 + x^2)}{x^2} f_{32}^\topoid{\thickbar{A}181}\\
       &- (1-2\epsilon)(1-3\epsilon)\epsilon^2\, \frac{ 2(1 + x^2)}{(1+x^2-x z)} f_{18}^\topoid{A166}
       - \epsilon^2\, \frac{ 5m^2(1 + x^2)}{4(1+x^2-x z)} f_{2}^\topoid{A134}
       - \epsilon^2\, \frac{ 9m^2 (1 + x^2)}{4x} f_{4}^\topoid{\thickbar{A}148}\\
       &- \epsilon^3\, \frac{ m^2 (1 + x^2) (1+x+x^2-xz)}{(1+x^2-xz) x} \left[ 3\, f_{17}^\topoid{\thickbar{A}149} + 2\, m^2 f_{36}^\topoid{\thickbar{A}182} \right]
       - \epsilon^4\, \frac{ 3m^2 (1 + x^2) (1 + z)} {(1+x^2-xz)} f_{35}^\topoid{\thickbar{A}182}\,,
\end{align*}
 \begin{align*}
 m_{74} =
       &- \epsilon^4\, \frac{m^4 (1 + x)^2 (1-2xz+x^2)}{x^2} f_{63}^\topoid{C126}
       + \epsilon^4\, \frac{m^6 (1 + x)^4 (1+x^2-xz)}{x^3} f_{73}^\topoid{C382}\\
       &- \epsilon^3\, \frac{ m^4 (1 + x)^2 (1 + x^2)}{x^2} \left[2 f_{30}^\topoid{A181} - f_{32}^\topoid{\thickbar{A}181} + \epsilon\, f_{74}^\topoid{C382} \right]
       + \epsilon^2\, \frac{ m^2(1+x)^2(1+x^2)(2-3xz+2x^2)}{4 (1+x^2-xz) x^2 z} \left[
           f_{1}^\topoid{A38} +  4\,\epsilon\, f_{19}^\topoid{A166} \right]\\
       &- \epsilon^2\, \frac{9(1 + x^2) m^2}{4x} \left[ 2\,f_{3}^\topoid{A148} - f_{4}^\topoid{\thickbar{A}148} \right]
       - (1-2\epsilon)(1-3\epsilon)\epsilon^2\,\frac{2 (1+x^2) (2-3xz+2x^2)}{(1+x^2-xz)xz} f_{18}^\topoid{A166}\\
       &- \epsilon^2\, \frac{5(1 + x^2) (2-3xz+2x^2) m^2}{4(1+x^2-xz) x z} f_{2}^\topoid{A134}
       - \epsilon^4\, \frac{6(1 + x^2)  (1+x+x^2-xz) m^2}{x^2 z} f_{33}^\topoid{A182}\\
       &+ \epsilon^3\, \frac{(1 + x^2) (1+x+x^2-xz) m^2}{(1+x^2-xz) x} \left[ 3\, f_{17}^\topoid{\thickbar{A}149} + 2\,m^2\,f_{36}^\topoid{\thickbar{A}182} \right]
       - \epsilon^3\, \frac{ (1 + x^2) (1 + z) m^2}{x z} \left[ 6 f_{16}^\topoid{A149} + 4\,m^2\,f_{34}^\topoid{A182} \right]\\
       &+ \epsilon^4\, \frac{3(1 + x^2) (1 + z) m^2}{(1+x^2-xz)}f_{35}^\topoid{\thickbar{A}182}\,,
\end{align*}
\begin{align*}
    m_{75} =
       &- \epsilon^2\, \frac{3 m^2 z}{4(1 + x)} f_{3}^\topoid{A148}
       + \epsilon^2\, \frac{3 m^2 (2+x+xz+2x^2-x^2z+x^3) }{4(1-2\epsilon)(1+x)x}f_{4}^\topoid{\thickbar{A}148}\\
       &+ \epsilon^2\, \frac{ m^2(1+x+4xz+x^2-x^2z+x^3)}{4(1-2\epsilon)(1+x) (1+x^2-x z)} f_{2}^\topoid{A134}
       - \epsilon^3\, \frac{(7+7x+3xz+3x^2-3x^2z+3x^3)}{(1+x)(1+x^2-xz)} f_{18}^\topoid{A166}\\
       &+ \epsilon^2\, \frac{(1+x+xz+x^2-x^2z+x^3)}{(1+x)(1+x^2-xz)} f_{18}^\topoid{A166}
       - \epsilon^4\, \frac{ 3 m^2(1+x+x^2-xz)}{(1 + x) x} f_{33}^\topoid{A182}\\
       &+ \epsilon^3\, \frac{ m^4(1+x+x^2-xz)  (3+x+xz+3x^2-x^2z+x^3)}{2(1-2\epsilon)(1+x)(1+x^2-xz) x}f_{36}^\topoid{\thickbar{A}182}\\
       &+ \epsilon^3\, \frac{ 3 m^2(1+x+x^2-xz)(1+x+xz+x^2-x^2z+x^3)}{2(1-2\epsilon)(1+x)(1+x^2-xz) x}f_{17}^\topoid{\thickbar{A}149} 
       - \epsilon^4\, \frac{ m^2(1-x^2)}{x} f_{75}^\topoid{C382}\\
       &- \epsilon^3\, \frac{ m^4(1 + z)}{(1 + x)} f_{34}^\topoid{A182}
       + \epsilon^4\, \frac{12}{(1+x^2-xz)} f_{18}^\topoid{A166}
       + \epsilon^4\, \frac{2 m^2(1 + x)}{(1-2\epsilon) x} f_{13}^\topoid{A53}
       - \epsilon^3\, \frac{9 m^2 }{2(1-2\epsilon)x} f_{4}^\topoid{\thickbar{A}148}\\
       &- \epsilon^4\, \frac{m^4 (1 + x) z}{x} f_{74}^\topoid{C382}
       + \epsilon^3\, \frac{ m^2 (1 + x) (2+x-xz+x^2)}{2(1-2\epsilon) (1+x^2-xz) x} f_{1}^\topoid{A38}\\
       &+ \epsilon^3\, \frac{ m^4(1 + x)(1+x+xz+x^2-x^2z+x^3)}{2(1-2\epsilon) x^2}   f_{32}^\topoid{\thickbar{A}181} 
       - \epsilon^4\, \frac{ 6m^2(1+x+x^2-xz)}{(1-2\epsilon)(1+x^2-xz) x} f_{17}^\topoid{\thickbar{A}149}\\
       &- \epsilon^4\, \frac{ 4 m^4 (1+x+x^2-xz)}{(1-2\epsilon)(1+x^2-xz) x} f_{36}^\topoid{\thickbar{A}182}
       - \epsilon^2\, \frac{ m^2(1+x)(1+x+x^2-x^2z+x^3)}{4(1-2\epsilon)(1+x^2-xz) x} f_{1}^\topoid{A38}\\
       &+ \epsilon^4\, \frac{ m^4(1+x)(1+z)}{x} f_{63}^\topoid{C126}
       - \epsilon^3\, \frac{ m^2(1+x)^2}{2(1-2\epsilon)x} f_{13}^\topoid{A53}
       + \epsilon^4\, \frac{ 2m^2(1 + x)^2}{(1-2\epsilon)(1+x^2-xz)x}f_{19}^\topoid{A166}\\
       &- \epsilon^4\, \frac{ 2m^4 (1 + x)^2}{(1-2\epsilon) x^2} f_{32}^\topoid{\thickbar{A}181}
       + \epsilon^4\, \frac{ m^4 (1 + x)^3}{2 x^2} f_{62}^\topoid{C318}
       - \epsilon^3\, \frac{ m^2(1+x)(1+x^2)}{(1-2\epsilon)(1+x^2-xz) x} f_{19}^\topoid{A166}\\
       &- \epsilon^5\, \frac{ 2m^2 (1+3x+2xz+x^3z-x^4)}{(1-2\epsilon)(1+x^2-xz) x} f_{35}^\topoid{\thickbar{A}182}
       + \epsilon^4\, \frac{ 3m^2 (1+x^2)(1+z)}{(1-2\epsilon)(1+x) (1+x^2-xz)} f_{35}^\topoid{\thickbar{A}182}\\
       &- \epsilon^3\, \frac{ m^2 (2+5x+3xz-3x^2)}{2(1-2\epsilon)(1+x)(1+x^2-xz)}  f_{2}^\topoid{A134}\,.
\end{align*}
}

We remark here that even if the formulas look in some cases rather 
cumbersome, they are always at most linear combinations of the starting basis $f_j$ with 
rational coefficients. Obviously, choosing differently this starting basis 
can simplify or even complicate substantially these relations.
On the other hand the main point of the derivation given in 
Section~\ref{sec:basis} is to show how, starting from a basis
whose differential equations fulfil
some initial requirements, a canonical basis (if existing!) 
can be built in an almost algorithmic way.

\bibliographystyle{JHEP}   
\bibliography{Biblio}     

\end{document}